\documentstyle[12pt]{article}
\def\be{\begin{equation}}
\def\en{\end{equation}}

\begin{document}
\baselineskip = 24pt
  \begin{flushright}
    \today
  \end{flushright}
\begin{center}

{\large \bf A new spherically symmetric general relativistic hydrodynamical
code.}

\vspace{2. cm}

{\bf Jos\'e V. Romero$^{1}$,
Jos\'e M$^{\underline{\mbox{a}}}$. Ib\'a\~{n}ez$^{2}$ \\
Jos\'e M$^{\underline{\mbox{a}}}$. Mart\'{\i} $^{2,3}$}, and
{\bf  Juan A. Miralles$^{2}$}\\
\vspace{0.5 cm}
\normalsize
\baselineskip 18pt
$^{1}$ Departamento de F\'{\i}sica Te\'{o}rica \\
Universidad de Valencia \\
46100 Burjassot (Valencia), Spain \\
$^{2}$ Departamento de Astronom\'{\i}a y Astrof\'{\i}sica \\
Universidad de Valencia \\
46100 Burjassot (Valencia), Spain \\
$^{3}$ Max-Planck-Institut f\"ur Astrophysik \\
Karl-Schwarzschild-str., 1, 8046 Garching bei M\"unchen, Germany

\end{center}

\normalsize

\begin{abstract}

\noindent

In this paper we present a full general relativistic one-dimensional
hydro-code which incorporates a modern high-resolution shock-capturing
algorithm, with an approximate Riemann solver,
for the correct modelling of formation and propagation of strong shocks.
The efficiency of this code in treating strong shocks is demonstrated by
some numerical experiments. The interest of this
technique in several astrophysical scenarios is discussed.
\end{abstract}


\normalsize
\newpage

\begin{center}
{\bf I. INTRODUCTION}
\end{center}

Since the pioneering work by May and White (1967) the use of general
relativistic and spherically symmetric hydro-codes has been restricted,
basically, to the field of stellar collapse and Supernovae. Currently, there
are several astrophysical scenarios for which the constraint of spherical
symmetry is still a good approximation and where general relativistic
hydrodynamical processes are involved: i) Gamma-ray bursters: Models for
explaining gamma-ray bursts compatible with the spatial distribution derived
from the BATSE experiment from the Compton Observatory are based on
relativistic fireballs originating from the sudden release of energy in small
regions (M\'esz\'aros et al. 1993, Piran et al. 1993). ii) Spherical accretion
onto compact objects: Theoretical studies of spectral properties of X-ray
radiation produced in atmospheres around an accreting neutron star have
particular importance given the observational capabilities of present
instrumentation on board satellites. In particular cases -- low magnetic
fields, for example -- the assumption of spherical symmetry is adequate. iii)
Stellar collapse: Different topics of current interest in the field of stellar
collapse and supernovae are, apart from the theoretical problem of black hole
formation (see, e.g., Baumgarte, Shapiro and Teukolsky 1994), the equation of
state for dense matter, the role of neutrinos, the influence of convective
motions, etc.. One of them is the correct modelling of formation and
propagation of the shock formed after bounce, the so-called {\it prompt phase}.
This question remains of crucial importance as an initial
mechanism leading, with the help of other processes involved in the different
versions of the {\it delayed mechanism} -- neutrino energy deposition (Bethe
and Wilson 1985), convective motions (Janka and M\"uller 1993, Herant et al.
1994) --, to the final success of the explosion.

The presence of strong shocks is a common feature in the above astrophysical
scenarios. Shocks are discontinuous solutions of the hydrodynamical equations
and are an important source of numerical problems and inaccuracies. The correct
modelling of strong shocks is one of the most delicate issues in -- both
Newtonian and relativistic -- hydrodynamical codes.

May and White's code was built up by using standard finite difference
techniques and incorporates an artificial viscosity term to damp down spurious
numerical oscillations around discontinuities. Up to date, codes based on the
original formulation of May and White and on later versions (e.g., Van Riper
1979) have been used in many supernovae calculations (see, e.g., the recent
paper by Swesty, Lattimer and Myra 1994 and references cited therein). The
Lagrangian character of May and White's code together with other theoretical
considerations concerning the particular coordinate gauge, has prevented its
extension to multidimensional calculations.

The first Eulerian code to solve the equations of relativistic hydrodynamics
comes from the work of Wilson (1972, 1979). The main ideas of Wilson's
procedure laid the foundations of several codes developed in the first half of
the eighties which have been applied to several astrophysical scenarios:
axisymmetric stellar collapse (Piran 1980, Stark and Piran 1987, Nakamura et
al. 1980, Nakamura 1981, Nakamura and Sato 1982, Evans 1986), accretion onto
compact objects (Hawley et al. 1984, Petrich et al. 1989) and numerical
cosmology (Centrella and Wilson 1984). All of these codes make use of a
combination of artificial viscosity and upwind techniques in order to obtain
numerical solutions of the relativistic hydrodynamical equations. On the other
hand, the equations are written as a set of advection equations, so
the terms containing derivatives (in space or time) of the pressure are
treated as source terms. This procedure breaks down (numerically) the {\it
conservative} character of the relativistic hydrodynamics system of equations
(see below).

Concerning the present work, during the last decade a number of new {\it
shock-capturing} finite difference ap\-pro\-xi\-ma\-tions have been constructed
and found to be very useful in the numerical simulation of classical
(Newtonian) fluid dynamics (see, e.g., LeVeque 1992). In addition to
conservation form, these schemes are usually constructed to have the following
properties: a) Stable and sharp discrete shock profiles. b) High accuracy in
smooth regions of the flow. Schemes with these characteristics are usually
known as {\it high-resolution shock-capturing} schemes (henceforth HRSC). They
avoid the use of artificial viscosity terms when treating discontinuities and,
after extensive experimentation, they appear to be a solid alternative to
classical methods with artificial viscosity.

Our spherically symmetric general relativistic hydro-code is the cumulative
result of our experience in HRSC schemes applied to Newtonian and
special-relativistic fluid dynamics (see Ib\'a\~{n}ez 1993, for a review). In
previous papers (Mart\'{\i} et al. 1991, Marquina et al. 1992) HRSC schemes
have been extended to solve the relativistic hydrodynamics system of equations
in one spatial dimension. The procedure relied on two main points: 1) To write
the equations of relativistic hydrodynamics as a system of {\it conservation
laws} and identify the suitable vector of unknowns. 2) An {\it approximate
Riemann solver} built up from the spectral decomposition of the Jacobian matrix
of the system at the boundaries of each numerical cell. Schneider et al. (1993)
have also explored the ultrarelativistic regime with a different Riemann
solver. The multidimensional extension of our special-relativistic hydro-code
can be found in Font et al. (1994). Similar results have been independently
derived by Eulderink (1993). Recently, Mart\'{\i} and M\"uller (1995) have
built up a relativistic version of the popular Newtonian Piecewise Parabolic
Method (PPM, Colella and Woodward 1984) and have carried out simulations of
relativistic jets with Lorentz factors greater than $20$ (Mart\'{\i} et al.
1995a).

Finally, note that in recent years, techniques other than the standard finite
difference ones have been applied to simulate relativistic flows. Gourgoulhon
(1991,1992) made use of the accuracy of {\it spectral methods} (Gottlieb and
Orszag 1977) to detect the zero value of the fundamental mode against radial
oscillations of a neutron star at the maximum of the mass-radius curve for a
given equation of state. Although the global error on the solution decreases
exponentially with the number of degrees of freedom, the handling of shock
waves by spectral methods remains a major problem due to the Gibbs phenomenon.
By combining moving grids and shock tracking techniques Bonazzola and Marck
(1991) have obtained promising results for one-dimensional flows. A different
numerical approach is provided by {\it smooth particle hydrodynamics} (Gingold
and Monaghan 1977, Lucy 1977), the first application to simulate relativistic
flows being made by Mann (1991). These results, although encouraging, are still
far from those obtained with HRSC methods.

General relativistic hydrodynamics links gravitational field with geometry,
hence, from the numerical point of view, new difficulties arise due to the
nonlienarities introduced by the geometrical terms. The system of equations is
coupled not only by Lorentz-like factors, as in special relativity (see Norman
and Winkler, 1986 for a discussion on the numerical difficulties), but also by
the different components of the metric tensor. In the present paper we describe
the features of a full general relativistic hydrodynamical code (some
preliminary results were presented in Ib\'a\~{n}ez et al. 1992) having the
following main properties: i) The code makes use of an approximate Riemann
solver which allows to capture shocks in a consistent way. It is a natural
extension to general relativistic fluid dynamics of modern HRSC schemes.  ii)
By using splitting techniques the code can be easily extended to the
multidimensional case once the behaviour of the characteristic fields is known.

The structure of this work is as follows: Section $\S II$ is devoted to display
the full equations of general relativity in a spherically symmetric space-time
as well as to the theoretical analysis of this system considered as a system of
conservation laws. The hydro-code is based on a numerical algorithm which is
explained in $\S III$. Some numerical tests and astrophysical applications are
shown in $\S IV$. Finally, a summary of our main conclusions is presented in
$\S V$.

\begin{center}
{\bf II. GENERAL RELATIVISTIC EQUATIONS IN SPHERICAL SYMMETRY
AS A SYSTEM OF CONSERVATION LAWS}
\end{center}

Let $\cal M$ be a general spacetime, described by the four
dimensional metric tensor $g_{\mu \nu}$.
According to the $\{3+1\}$ formalism, the metric is split into the
objects $\alpha$ ({\it lapse}), $\beta^{i}$ ({\it shift}) and
$\gamma_{ij}$, keeping the line element in the form:
\begin{equation}
ds^{2} = -(\alpha^{2}-\beta_{i}\beta^{i}) dt^{2}+
2 \beta_{i} dx^{i} dt + \gamma_{ij} dx^{i}dx^{j}
\end{equation}
\noindent
where Greek (Latin) indices run through all (spatial)
coordinates.

In the case of spherically symmetric space-times the general
relativistic equations can be given in a simple way which looks
like the Newtonian hydrodynamics. To this aim the choice of
coordinates is crucial.
Schwarzschild-type coordinates (Bondi 1964) allow a simple extension
of the Eulerian Newtonian hydrodynamics to the Einstenian one.
In terms of slicing of space-time, Schwarzschild-type coordinates are
the realization of a {\it polar time slicing}, a {\it radial gauge}
(see Gourgoulhon 1991), and the generalization of the Schwarzschild
coordinates, in terms of which the vacuum and static space-time
is described.

The choice of the radial gauge leads to the following expression for
the 3-metric
\begin{eqnarray}
\gamma_{ij} = {\rm diag}(X^{2}, r^{2}, r^{2}\sin^{2}\theta) \,\,\,\,.
\label{gammaij}
\end{eqnarray}

The choice of the polar slicing condition, in spherical symmetry
and the radial gauge implies a zero-shift vector ($\beta^i = 0,  \forall i$).
In the following, we
will use the acronym RGPS (radial gauge and polar slicing)
for the particular spacetime in which we are interested:
\begin{equation}
ds^{2} =  - \alpha^2 dt^{2} + X^2 dr^{2} +
r^2 (d \theta^2 + sin^2 \theta d\phi^2) \,\,.
\label{RGPS}
\end{equation}

By analogy with the well-known Schwarzschild solution for
vacuum, we can define the functions $\Phi (r,t)$ and $m(r,t)$ by
\begin{eqnarray}
X(r,t) = \left(1-\frac{2m(r,t)}{r} \right)^{-1/2} \,\,\,\,\,\,\,\, ,
\,\,\,\,\,\,\,\, \alpha(r,t) = \exp\{\Phi(r,t)\} \,\,\,\,.
\end{eqnarray}

The equations describing the evolution of matter
are the expression of
the {\it local conservation} of baryon number
\begin{equation}
\nabla_{\mu}J^{\mu} = 0   \,\,\,\,,
\label{J}
\end{equation}
\noindent
and the {\it local conservation} of energy-momentum
\begin{equation}
\nabla_{\mu}T^{\mu\nu} = 0  \,\,\,\,,
\label{T}
\end{equation}

The current $J^{\mu}$ and the energy-momentum tensor $T^{\mu\nu}$
are
\begin{equation}
J^{\mu} = \rho u^{\mu} \,\,\,\,,
\end{equation}
\begin{equation}
T_{\mu\nu} = \rho h u_{\mu}u_{\nu} + p g_{\mu \nu}  \,\,\,\,,
\end{equation}
\noindent
where $\nabla_{\mu}$ is the covariant derivative,
$u^{\mu}$ the four-velocity of the fluid,
$\rho$ the rest-mass density, $p$ the pressure, and
$h$ the specific enthalpy defined by $h = 1 + \varepsilon + p/\rho$, where
$\varepsilon$ is the specific internal energy.

The expression chosen for the energy-momentum tensor is that of a perfect fluid
and we therefore ignor effects due to heat conduction or viscous interactions.

Let us introduce the {\it physical velocity},
$v$, defined by $v = X u^{r}/ \alpha u^{t}$. This quantity represents
the fluid velocity relative to an
observer at rest in the coordinate frame. The Lorentz-like factor
$W = \alpha u^{t}$, satisfies the familiar relation with $v$,
$W = (1-v^{2})^{-1/2}$.

With the above coordinate conditions and the following set
of unknown variables
\begin{eqnarray}
D & = & \alpha X J^t = X \rho W  \,\,\,\,, \nonumber \\
S & = & \alpha T^{tr} = \rho hW^{2}v \,\,\,\,, \nonumber \\
\tau & = & \alpha^{2} T^{tt} - D  =  \rho hW^{2} - p - D \,\,\,\,,
\end{eqnarray}
the general relativistic equations can be written
as a system of conservation laws (with sources):
\begin{eqnarray}
\frac{\partial {\bf u}}{\partial t}+
\frac{1}{r^2} \frac{\partial \left(\displaystyle{
\frac{r^2 \alpha}{X}}{\bf f(u) }\right)}{\partial r}
= {\bf s(u)}  \,\,\,\,,
\label{sys1}
\end{eqnarray}
\noindent
where
\begin{eqnarray}
{\bf u} = {\it (D, S, \tau)} \,\,\,\,,
\end{eqnarray}
\noindent
is the {\it vector of unknowns} which define the state of the
system. The {\it fluxes}, ${\bf f}$, are defined as
\begin{eqnarray}
{\bf f} = \left( D v , S v + p, S - D v \right) \,\,\,\,,
\end{eqnarray}
\noindent
and the {\it source terms} (free of derivatives of hydrodynamic
quantities) are
\begin{eqnarray}
{\bf s(u)}  =  \left(0, (Sv - \tau - D )(8 \alpha X \pi rp + \alpha X
\frac{m}{r^2}) + \alpha X p \frac{m}{r^2} + \frac{2 \alpha p}{X r}
                      ,0 \right) \,\,\,\,.
\end{eqnarray}

The above system of equations is closed by an
equation of state (EOS), that we will assume of the form
\begin{eqnarray}
p = p(\rho,\epsilon) \,\,\,\,,
\end{eqnarray}
\noindent
and the Einstein equations which furnish conditions on
the quantities
$m(r,t)$ and $\Phi (r,t)$ (see, Gourgoulhon 1991):
\begin{eqnarray}
\frac{\partial m}{\partial r} = 4 \pi r^{2} (\tau + D) \,\,\,\,,
\label{mg}
\end{eqnarray}
\begin{eqnarray}
\frac{\partial \Phi}{\partial r} =
X^{2} \left( \frac{m}{r^{2}}+
4 \pi r (p + S v) \right) \,\,\,\,,
\label{phi}
\end{eqnarray}

Indeed, by analogy with the static case, we can distinguish (for
a given spherical surface having an area $4 \pi r^{2}$) between the
enclosed {\it gravitational mass} $m$ defined by (\ref{mg}) and the
enclosed {\it baryonic mass} $m_A$ defined by
\begin{eqnarray}
\frac{\partial m_A}{\partial r} = 4 \pi r^{2} D \,\,\,\,.
\label{ma}
\end{eqnarray}

Equations (\ref{mg}) and (\ref{ma}) are integrated, at
each time step, between $r=0$ and $r=R(t)$ (the radius of the star) with
the following boundary conditions:
$m(r=0) = 0$, $m_A(r=0) = 0$,
$m(R(t)) = M$ (the total gravitational mass),
$m_A(R(t)) = M_A$ (the total baryonic mass). The gravitational potential
$\Phi$, given by equation (\ref{phi}), is defined with the exception of an
aditive arbitrary constant and
is matched at the surface with the exterior Schwarzschild's solution,
i.e., $\Phi(R(t)) = (1/2) {\rm ln}(1-2M/R(t))$. The integration
of $\Phi$ starts with a zero value at the center and
by comparing, at the surface, the integrated value of $\Phi$ with
the matching condition it is possible to obtain the arbitrary aditive
constant.

It is worth pointing out that
the system of equations (\ref{sys1})
displays a very important feature, that is,
the conservation of baryonic mass and energy.
In effect, the first and third equations of system
(\ref{sys1}) lead (when acting on integral quantities) to the following
relations:
\begin{eqnarray}
\frac{\partial M_A}{\partial t} & = & - \frac{\alpha}{X} Dv \mid_{R(t)}
                                        \,\,\,\,,\nonumber \\
\frac{\partial E_b}{\partial t} & = & - \frac{\alpha}{X} (S-D)v \mid_{R(t)}
\,\,\,\,.
\end{eqnarray}
\noindent
The quantity $E_b = M - M_A$, on the analogy of the static case, can be
considered as the binding energy of the star. The symbol $\mid_{R(t)}$ means
that the quantities on the right hand side of these equations are evaluated at
the surface. Hence, if the boundary conditions at the surface are zero (i.e.,
$\rho$ = $\epsilon$ = 0, or $v=0$), the conservation of $M_A$ and $E_b$ (or
$M$) is strictly satisfied.

A numerical algorithm written in conservation form, like the one
used in the present paper (see below),
preserves numerically this important property of the system.
Hence, the correct choice of the
vector of unknowns $\bf u$,
and consequently, source terms $\bf s(u)$ is of crucial importance.

 From the {\it conserved} quantities ${\cal C} = \{D, S, \tau \}$ we must
obtain
the set of {\it primitive variables} $\wp = \{\rho, \it v, \epsilon \}$, at
each
time step, by solving an implicit equation in pressure. A one-dimensional
Newton-Raphson routine suffices to obtain $\wp$
(see Mart\'{\i} \& M\"uller 1995, for details).
In the Newtonian limit,
the set of new variables ${\cal C} = \{D, S, \tau \}$ tends to the set $\{\rho,
\rho v, e \}$, with $e = \rho \epsilon + (1/2)\rho v^{2} - m/r$,
i.e., the density, momentum density
and total energy density, respectively.

The hyperbolic character of the relativistic hydrodynamic system of equations
has been the subject of study of many authors
(see Anile, 1989 and references therein).

As we will show in the next section, Godunov-type methods, or modern HRSC
schemes, incorporate the resolution of {\it local Riemann problems},
initial value problems for system (\ref{sys1}) with discontinuous data.
In order to extend HRSC to our problem is crucial to know
the spectral decomposition of the Jacobian matrix
${\bf \cal B}$({\bf u}) of the system (\ref{sys1}):
\begin{equation}
{\bf \cal B} = \frac{\partial{\bf f}({\bf u})}
{\partial\bf u \rm}  \,\,\,\,.
\label{B}
\end{equation}

Following a procedure similar to the one described in Font et al.
(1994), we have derived
the eigenvalues and right-eigenvectors of ${\bf \cal B}$({\bf u}).

The {\it eigenvalues}, the characteristic speeds associated with
{\it material waves} and {\it acoustic waves}, are respectively,
\begin{eqnarray}
\lambda_{0} = v \,\,\,\,,
\end{eqnarray}
\noindent
and
\begin{eqnarray}
\lambda_{\pm} = \frac{v \pm c_{s}}{1 \pm v c_s} \,\,\,\,,
\end{eqnarray}
\noindent
where $c_{s}$ is the {\it local sound velocity} which satisfies
\begin{equation}
h c_{s}^{2} = \chi + (p/\rho^{2})\kappa  \,\,\,\,,
\label{cs}
\end{equation}
\noindent
with $\chi = (\partial p / \partial \rho)_{\epsilon}$ and $\kappa =
(\partial p / \partial \epsilon)_{\rho}$.

These eigenvalues are
the natural extension to our generic
space-time of the corresponding characteristic speeds well-known in
Minkowski space-time or in Newtonian fluid dynamics.
Let us point out that they seem like in Minkowski space-time due to
our definitions of velocity and flux.

The {\it right-eigenvectors} are:
\begin{eqnarray}
{{\bf r}}_{0}   & = & \left(\frac{X \widetilde{\kappa}}
{h W (\widetilde{\kappa} - c_s^2)}, v, 1- r^{(1)}_{0}    \right) \,\,\,\,,\\
{{\bf r}}_{\pm} & = & \left( \frac{X (1- \lambda_{\pm}  v)}
{h W(1-v^2)}, \lambda_{\pm} , 1- r^{(1)}_{\pm}     \right) \,\,\,\,,
\end{eqnarray}
\noindent
where $\widetilde{\kappa} = \kappa / \rho$,
and $r^{(1)}_{0,\pm}$ is the first component of the corresponding
eigenvector.

The above system (\ref{sys1}) is strictly hyperbolic since the Jacobian
matrix ${\bf \cal B}$({\bf u})
has real and distinct eigenvalues.

\begin{center}
{\bf III. OUR MODERN HIGH-RESOLUTION SHOCK-CAPTURING SCHEME}
\end{center}

In order to exploit numerically the conservative character of the
system (\ref{sys1})
we have written it as a hyperbolic system of conservation laws
(see Lax 1972, for a mathematical analysis). Let us
summarize in this Section the main features of our algorithm
(see also Mart\'{\i} et al. 1991).

At each time level, the data are the {\it cell averages} of the
conserved quantities
\begin{equation}
\bar{\bf u}^{n}_{j} = \frac {1}{r^2_j \Delta r_{j}}\int_{r_{j-{1\over 2}}}^
{r_{j+{1\over 2}}}
{\bf u}(r,t^{n}) r^2  dr   \,\,\,\,.
\label{aver}
\end{equation}

The data are advanced in time according to a version of the
{\it method of lines}:
\begin{eqnarray}
\frac{d \bar{\bf u}_j(t) }{d t}=
-\frac{A_{j+{1\over 2}} \hat{\bf f}_{j+{1\over 2}}
- A_{j-{1\over 2}} \hat{\bf f}_{j-{1\over 2}}}
{r^2_j \Delta r_j} + \bar{\bf s}_j \,\,\,\,,
\label{sys}
\end{eqnarray}
\noindent
where
\begin{eqnarray}
\hat{\bf f}_{j+{1\over 2}} =
\hat{\bf f}({\bf u}_{j-1},{\bf u}_{j},
{\bf u}_{j+1}) \,\,\,\,,
\end{eqnarray}
\noindent
is a consistent {\it numerical flux} vector, i.e.,
$\hat{\bf f}({\bf u},{\bf u},{\bf u}) = {\bf f}({\bf u})$.
The quantity $A_{j+{1\over 2}}$
is a combination of the geometrical factors
\begin{eqnarray}
A_{j+{1\over 2}} =
\left(\displaystyle{\frac{r^2 \alpha}{X}}
\right)_{j+{1\over 2}} \,\,\,\,,
\end{eqnarray}
\noindent
evaluated at the interface $j+{1\over 2}$,
and $\bar{\bf s}_j$ is the cell average of the source terms calculated
according to (\ref{aver}).

Once the procedure to evaluate $\hat{\bf f}_{j+{1\over 2}}$ is known , then
system (\ref{sys}) can be integrated in time by using a suitable ordinary
dif\-fer\-ential equation solver. We have made use of a standard
predictor-corrector method. The value of the timestep is constrained by
the Courant condition.

A {\it reconstruction procedure} of the solution at the time level $t^n$ from
its cell averages allows us to define local Riemann problems at each interface
$j+{1\over 2}$. We have used a monotonicity preserving linear reconstruction
of the
primitive variables using the {\it minmod} function as a 'slope limiter' (Van
Leer, 1979). Accordingly, the corresponding values of ${\bf u}_{j+{1\over 2}}$
at the
interface $j+{1\over 2}$are:
\begin{equation}
{\bf u}^{L}_{j+{1\over 2}} = \bar{\bf u}^{n}_{j} +
{\bf S}^n_j (r_{j+{1\over 2}} - r_j)
\,\,\,\,,
\end{equation}
\begin{equation}
{\bf u}^{R}_{j+{1\over 2}} = \bar{\bf u}^{n}_{j+1} +
{\bf S}^n_{j+1} (r_{j+{1\over 2}} - r_{j+1}) \,\,\,\,,
\end{equation}
\noindent
where ${\bf S}^n_j$ is a {\it slope limiter} defined by
\begin{equation}
{\bf S}^n_j = {\rm minmod} \left( \frac{\bar{\bf u}^{n}_{j+1} -
\bar{\bf u}^{n}_{j}}{r_{j+1}-r_j},
\frac{\bar{\bf u}^{n}_{j} - \bar{\bf u}^{n}_{j-1}}
{r_{j}-r_{j-1}} \right)  \,\,\,\,,
\end{equation}
and the $minmod$ function makes a choice of the slope which is minimum
or takes a zero when they have different signs:

\[
{\rm minmod (a,b)} = \left\{ \begin{array}{ll}
                        a & \mbox{if $\mid a \mid < \mid b \mid, ab>0$} \\
                        b & \mbox{if $\mid a \mid > \mid b \mid, ab>0$} \\
                        0 & \mbox{if $ab\le0$}
                        \end{array}
                  \right. \,\,\,\,. \]

At each cell interface the
$i^{th}$ component of the {\it numerical flux} is computed as follows
\begin{eqnarray}
\widehat {f}_{j+{1\over 2}}^{(i)}
 = \frac{1}{2}
\left( f^{(i)}({\bf u}_{j+{1\over 2}}^{L})  +
f^{(i)}({\bf u}_{j+{1\over 2}}^{R}) -
\sum_{\alpha = 0,\pm} \mid \widetilde{\lambda}_{\alpha}\mid
\Delta \widetilde {\omega}_{\alpha}
\widetilde {r}_{\alpha}^{(i)}\right) \,\,\,\,,
\end{eqnarray}
where L and R stand for the left and right states at a given
interface $j+{1\over 2}$,
$\widetilde {\lambda}_{\alpha}$ and
$\widetilde {r}_{\alpha}^{(i)}$
($\alpha=0,\pm$) are respectively, the eigenvalues and the
$i^{th}$-component of the $\alpha$-right eigenvector of the
Jacobian matrix
\begin{eqnarray}
{\bf \cal B}_{j+{1\over 2}} = \left(\frac{\partial {\bf f(u)}}
{\partial{\bf u}} \right)_{{\bf u}={\frac{1}{2}}({\bf u}^{R}_{j+{1\over 2}} +
{\bf u}^{L}_{j+{1\over 2}})} \,\,\,\,.
\end{eqnarray}

The quantities $\Delta \widetilde {\bf \omega}_{\alpha}$,
the jumps in the local characteristic variables across each cell
interface, are obtained from
\begin{eqnarray}
{\bf u}^{R}_{j+{1\over 2}} - {\bf u}^{L}_{j+{1\over 2}} = \sum_{\alpha = 0,
\pm}
\Delta\widetilde{\bf \omega}_{\alpha}
\widetilde {\bf r}_{\alpha} \,\,\,\,.
\end{eqnarray}
\noindent
$\widetilde {\lambda}_{\alpha}$, $\widetilde {\bf r}_{\alpha}$ and $\Delta
\widetilde {\bf \omega}_{\alpha}$, as functions of $\bf u$, are evaluated at
each interface and, therefore, they depend on the particular values ${\bf
u}^{L}_{j+{1\over 2}}$ and ${\bf u}^{R}_{j+{1\over 2}}$.

With the above ingredients, our algorithm
is conservative, upwind, and due to the
particular cell-reconstruction, monotone. It
is globally second order accurate, although this statement
is only well-founded for scalar equations, equally spaced
grids, and the smooth part of the flow.

Finally, let us comment on one of the features of our code, its
{\it modularity}. It has
been constructed to allow for easy substitution of different
Riemann solvers and different cell reconstructions. It also works
in Newtonian and special relativistic hydrodynamics.

\begin{center}
{\bf IV. NUMERICAL EXPERIMENTS AND ASTROPHYSICAL APPLICATIONS}
\end{center}

Modern numerical codes must be, and in fact are, checked against
well-known analytical solutions before their exploitation in applications. A
long
battery of test-beds addressed to check hydrodynamical codes exists. The most
popular ones are the {\it standard shock problems} in different
symmetries, planar (shock-tube tests), cylindrical or spherical. Their
corresponding analytical solutions (in Newtonian fluid dynamics) can be
found in standard textbooks (see,e.g., Courant and Friedrichs 1948). The
one-dimensional solution of the Riemann problem for relativistic hydrodynamics
has been derived recently by Mart\'{\i} and M\"uller (1994) allowing for the
analytical solution of any initial value problem. In previous papers we have
carried out numerical experiments in {\it planar symmetry} with the following
standard shock-tube problems: Sod's test, blast wave and shock reflection
tests, in both Newtonian and special-relativistic hydrodynamics (see Mart\'{\i}
et al. 1991, and Marquina et al. 1992). Here, in Section $\S IV.1$, we have
concentrated on the {\it relativistic spherical shock reflection}
problem, which allows to check special relativistic dynamics in a spherically
symmetric geometry.

A second test
(see Section $\S IV.2$) is the {\it spherical accretion} onto a
self-gravitating object, including general relativistic effects. This test has
the advantage of allowing one to check stationary solutions of general
relativistic
hydrodynamics on a fixed background.

A third test (see Section $\S IV.3$) underwent by our hydro-code is the
detection of the {\it zeros of the fundamental mode} against radial
oscillations of a spherical equilibrium configuration, which
checks the
stability of equilibrium solutions.

In Section $\S IV.4$, we have analyzed Oppenheimer-Snyder
collapse. This is one of the standard tests for spherically symmetric, fully
relativistic codes, which allows one to check general relativistic dynamics.

With the above selection of tests we span a broad range of
different features: special relativistic effects, geometrical effects, general
relativistic effects in a background and fully general relativistic dynamics.

Finally,
in Section $\S IV.5$ we have carried out an analysis of the dynamics of
the general relativistic
collapse of compact objects.

\begin{center}
{\bf IV.1 Relativistic spherical shock reflection}
\end{center}

Noh (1987) has exhaustively analized the {\it spherical shock reflection}
problem in the Newtonian dynamics of ideal gases ($p = (\Gamma -1) \rho
\varepsilon $). Noh's paper is, in fact, devoted to the study of intrinsic
errors which appear when using artificial viscosity techniques in spherically
symmetric applications. We have considered the
relativistic version of this numerical experiment. The relativistic spherical
shock reflection is a severe test due to the difficulties connected with the
geometry and, in its relativistic version, with strong nonlinearities
induced by the Lorentz factor in the ultrarelativistic regime.

Analytical solutions for the relativistic shock reflection with planar,
cylindrical and spherical symmetry can be found in Mart\'{\i} et al. (1995b).
The initial data consists of an inflowing cold (i.e.,
$\varepsilon = 0$) gas with coordinate velocity $v$ and corresponding Lorentz
factor $W$. In the spherical case, the flow converges
towards the center and its reflection causes the compression and
heating of the gas as it converts its momentum into internal energy. A shock
is formed which starts to propagate through the inflowing gas. Behind the
shock the gas is at rest and, according to the conservation of energy
accross the shock, has a specific internal energy given by
\begin{equation}
\varepsilon^{+} = W - 1.
\end{equation}
The jump in density at the shock is defined by $\Delta \rho = \rho^+/\rho^-$,
where $\rho^+$ and $\rho^-$ are the postshock and preshock values,
respectively.
For a relativistic strong shock, like the one developed herein,
this jump satisfies
\begin{equation}
\Delta \rho  =   \frac{\Gamma W + 1}{\Gamma - 1},
\end{equation}
with a shock velocity given by
\begin{equation}
v_s = \frac{(\Gamma -1) W |v|}{W+1}.
\end{equation}

In the nonshocked part of the flow (i.e., $r \in ]v_s t, \infty [$), the
proper rest-mass density distribution is given by
\begin{equation}
\rho = \left( 1 + \frac{vt}{r}\right)^2 \rho_0
\end{equation}
where $\rho_0$ is the rest-mass density at infinity.

It is worthwhile to point out one relevant difference between
classical and relativistic shocks. In the relativistic case, the
jump in density is unbounded, whereas in the present test,
the ultrarelativistic regime leads to the following asymptotic
relations ($\rho_0 =1$),
\begin{eqnarray}
\Delta \rho & \rightarrow &  \frac{\Gamma}{\Gamma - 1} W \,\,\,\,,
                                        \nonumber \\
\rho^+ & \rightarrow & \left( \frac{\Gamma}{\Gamma - 1} \right)^3 W
                                      \,\,\,\,,  \nonumber \\
v_s & \rightarrow & (\Gamma -1) \,\,\,\,.
\end{eqnarray}

The initial data for this problem are defined in a unit
sphere ($0 \leq r \leq 1$) and for an ideal gas with $\Gamma=4/3$;
$v(r,0)=-v_0 (v_0 > 0)$, $\rho (r,0)=\rho_0=1$,
$\varepsilon (r,0)=\varepsilon_0=0$. For numerical
reasons, the initial value of the specific internal energy of the inflow gas
was set to a small nonzero value $\varepsilon_0=10^{-6} W_0$. The boundary
conditions are $v(1,t)=-v_0$ and the exact solution of $\rho$ at $r=1$
for all time.

We switched off the gravitational terms in our code and transformed
it into a purely special relativistic hydrodynamical code, and
have used a grid with 200
equidistant zones. The results displayed in Table 1 and Fig.1, correspond to
a particular instant of the evolution after 1000 timesteps. In order to
emphasize the relativistic effects, we have
done the calculation for a large sample of different initial
inflow velocities $v_0$.
Several
conclusions can be established from the data contained in Table 1: i)
The postshock density increases with the initial inflow velocity and
is unbounded. This is a typical relativistic effect.
ii) The ratio $\rho^+
/W_0$ tends to the asymptotic value of 64 when $v_0 \longrightarrow 1$,
which is consistent with the above asymptotic relations. iii) The
maximum of the
relative errors for the postshock density, $\varepsilon^{\rm max}_r$, and the
mean relative error of the same quantity, $\bar{\varepsilon}_r$, are
independent of the initial Lorentz factor and converge to 14\% and 2\%,
respectively. This feature of relative errors has been pointed out by
Mart\'{\i} and M\"uller (1995) in discussions of planar shock wall test
experiment with their relativistic version of PPM. In the
mean relative error we have not considered the zone next to the center, which
always dominates the maximum error, due to the well-known effect of numerical
overheating. Figure 1 shows our numerical results compared with the analytical
ones.  The overheating phenomena is less severe by far than the one found by
Noh (1987) in his analysis of the corresponding Newtonian problem. Noh (1987)
reported errors of the order of 1000\% in some of the experiments (in
Newtonian hydrodynamics) with artificial viscosity techniques. A
maximum relative error of 14\% in the postshock density (at the center)
is comparable to
the one obtained with the Newtonian PPM
reported by Noh (1987, see Fig.~24).

\begin{center}
{\bf IV.2 Spherical accretion onto a black hole}
\end{center}

We have tried to reproduce some of the stationary
solutions of the {\it  spherical accretion onto a black hole} in two cases: i)
{\it dust} accreting onto a Schwarzschild black hole (Hawley et al. 1984) and
ii) an {\it ideal gas} accreting onto a Schwarzschild black hole (Michel,
1972).

In these tests, the gravitational field is kept fixed. The initial conditions
are those of a vacuum, i.e., density is zero everywhere except at the outer
boundary where a gas is being continously injected with a velocity and
density given by the exact solution. We have run our code using two different
grids of 50 and 100 points, which span the interval $1.05 \leq r/2M \leq
 10.0$. Outflow boundary conditions have been taken at the inner boundary
$r=2.1M$.

The analytical solution of a spherical geodesic ({\it presureless}) flow
accreting onto a black hole can be found in Hawley et al. (1984). Table
2 summarizes relative errors in the three fundamental variables when
compared with the analytical solution at a time of 180 $M$ (when the stationary
solution has been reached). Discrepancies with the exact solution amount to
less
than $2\%$ for the maximum of the relative errors and less than $1\%$ for the
mean relative error. As it should happen these errors decrease under grid
refinement (compare first and second rows in Table 2).

The stationary solution of an {\it ideal gas} accreting spherically  onto a
Schwarzschild black hole was derived by Michel (1972). A critical point
exists, as in the Newtonian description, which can create numerical
difficulties. We have considered only solutions of accreting gas having a
critical point far outside of our computational domain. Table 2 summarizes the
relative errors in the three fundamental variables when compared with the
analytical solution at a time of 720 $M$, when the stationary solution has been
reached. Discrepancies with the exact solution amount to less than $3\%$ for
the maximum of the relative errors (less than $1\%$ for the mean relative
error). As before, these errors decrease when the grid is refined.
Figures 2 and 3 show relativistic rest mass
density and the velocity profiles. Three snapshots of quantity $D$ are plotted
in Fig. 2 and show that the numerical solution converges to the
stationary analytical one. The velocity converges so fast that it's
not possible to distinguish the same snapshots in Fig. 3.

As can be deduced from the behaviour of the mean errors under grid refinement
(see Table 2), our code is second order accurate for continuous solutions.

\vspace{1.cm}

\begin{center}

{\bf IV.3 Detection of the zero value for the fundamental mode against radial
oscillations of a compact object}

\end{center}

According to the {\it static stability criterion} (Harrison et al. 1965)
it is possible to
establish a correlation between the critical points
of a curve (in a "gravitational mass versus radius"  $M-R$ diagram)
built up
from equilibrium configurations obeying a given cold EOS and
the onset of instability of the zero-frequency  mode
against radial oscillations. This criterion avoids solving the
Sturm-Liouville equation governing radial perturbations of
an equilibrium model (see
Shapiro and Teukolsky 1983, for a general
discussion, or Mart\'{\i} et al. 1988, for details on the
numerical solution)
A hydro-code should allow for the dynamical study of perturbations
over an equilibrium configuration which can be compared
with predictions of the static stability criterion.
The dynamical detection of a
zero in the fundamental mode has been used as a test-bed of
hydro-codes by several authors (see, e.g., Ib\'a\~{n}ez 1984 and
Gourgoulhon 1991). The most accurate results
known to the authors are
those obtained by Gourgoulhon (1991) using pseudospectral techniques.

In our case we have focussed on a particular critical
point, the maximum of the $M-R$ curves corresponding to
two kind of compact objects (white dwarfs and neutron stars)
which obey suitable EOS (see below) and such that their
stability properties have been exhaustively studied (see references
below). That maximum separates the curve in two branches,
being the stable (unstable) one the corresponding to models having
central denstities lower (higher) than the critical one.

The calculations of
this Section as well as those of Section $\S IV.5$ have been carried out with a
numerical grid (which is Eulerian, i.e., fixed) built up in such a way that the
radius of the initial model is partitioned into N zones, distributed in
geometric progression in order to have finer resolution near the center. In our
simulations we have taken N=330 and the surface radius $R(t)$ is given by the
condition $\rho (R(t)) = \rho (R(t=0)) \approx 10^{5}$g/cm$^{3}$. Hence, the
data relative to the initial stellar model are given in the entire grid. As
time goes on and, according to the dynamics of each problem, the number of
cells covering the star decreases. During the contraction phase of our
numerical experiments in this Section ($\S IV.3$), or during the infall epoch
of the applications in Section $\S IV.5$, the surface is moved in accordance
with the above prescription ($\rho (R(t) = \rho (R(t=0))$) and we eliminate
those cells remaining outside R(t) (in practice, we impose over these cells the
vacuum conditions). The important point is that in none of these applications
the number of numerical cells used for describing the star are less than 200
(for N=330) keeping the numerical resolution in a good level for an accurate
calculation. An analogous procedure is established for the expanding phase.
Other prescriptions for the surface condition such as, e.g., a combination of
moving grids with the integration of the velocity at the surface would be
interesting to be studied.

\vspace{1.cm}

{\it IV.3.1 Detection of a zero in the fundamental mode against radial
oscillations of a white dwarf}

We have considered a sample of initial models which are white dwarf-like
configurations, obtained by solving the stellar structure equations with the
equation of state of an ideal gas of electrons including Coulombian corrections
due
to the ions (Ib\'a\~{n}ez, 1984) and with a homogeneous chemical
composition (Carbon).
The maximum gravitational mass model
has a value of $1.3862 M_{\odot}$.

We have found a change in the
stability behaviour at some point between $1.3847 M_{\odot}$ which is a stable
model and $1.3853 M_{\odot}$ which is an unstable one and collapses in a time
which is an order of magnitude higher than the characteristic dynamical one.
The corresponding values of the initial central densities are 1.3 and 1.5,
respectively, in units of $10^{10}$gcm$^{-3}$.

Figures 4 and 5 show the
velocity field as a function of time and radius for, respectively, the stable
and unstable models. We point out the different temporal scales
involved and the different interval of values spanned by the velocity in each
case.

We have also run simulations with a grid of 180 points and taken initial
models of constant density. These models are not equilibrium models and,
consequently, they start to collapse until some new equilibrium
configuration is reached. We find a change in the stability of
the models at some point between $1.37 M_{\odot}$
and $1.38 M_{\odot}$ which is unstable.

\vspace{1.cm}

{\it IV.3.2 Detection of a zero in the fundamental mode against radial
oscillations of a neutron star}

We have considered a family of neutron stars
obeying the EOS derived by Diaz-Alonso (1985)
in a field theoretical model approach to neutron matter at zero
and finite temperature (Diaz-Alonso et al 1989).

The model describes a many-body system of mutually
interacting nucleons and mesons: a fictitious $\sigma$ particle, pions,
$\rho$ and $\omega$ mesons. The interaction is given by a relativistic
Lagrangian containing the free Dirac Lagrangian for the nucleons,
the Lagrangians for the $\sigma$ particle, pion and
$\omega$ and $\rho$ mesons.
The interaction is described by two pieces: one
containing the meson-nucleon interaction in the form of Yukawa-like
couplings, and one which describes the meson-meson interaction.
The attractive part of the nuclear interaction is provided by the $\sigma$
and pion exchange. The $\omega$ and $\rho$ mesons give
rise to a repulsive short-range interaction, which is charge-dependent in the
latter case, in such a way that the EOS behaviour is different for nuclear
symmetric matter as for pure neutron matter. Details on the solution of the
model can be found in the references above.

We will focuss on the EOS II in Diaz-Alonso (1985) which
was derived fitting the free
parameters in the Lagrangian in such a way that,
for symmetric nuclear matter, the model saturates at a density
$\rho_n \ =\ 2.837 \times
\ 10^{14}$ $gcm^{-3}$. The corresponding binding energy is $-15.68$ MeV,
the symmetry energy $33.17$ MeV, and the nuclear incompressibility is
$K \ =\ 225$ MeV.

The equilibrium configurations obeying this EOS at zero temperature
have been taken as initial models in our hydro-code. Their
macroscopic properties were
exhaustively studied in Diaz-Alonso and Ib\'a\~{n}ez (1985) for the
cold version of EOS II.
The maximum gravitational mass
is $1.94556 M_{\odot}$
which corresponds a central energy density of
$2.48 \times 10^{15}$gcm$^{-3}$.
The reader interested on the properties of stability against radial
oscillations and the equilibrium features of slowly rotating stars
obeying the hot version of that EOS can address, respectively,
to Mart\'{\i} et al. (1988) and
Romero et al. (1992).

We have generated a perturbation of the equilibrium models by imposing
an initial velocity profile according to the law $v(r) = -v_o (r/R)$.
This procedure has the advantage of allowing a clear distinction between
the dynamics of the models, erasing round-off errors induced in the
numerical construction of the initial model (for the hydro-code) from
the model generated by solving the structure equations.
As before, the value of $R(t)$
is obtained from the condition: $\rho (R(t)) = \rho (R(t=0))$.

Taking a value for $v_o$ of $10^{-3}$ we have found a change
in the stability behaviour between the model of
$1.94532 M_{\odot}$ (central density:
$2.55 \times 10^{15}$gcm$^{-3}$), which is stable, and the model
$1.94518 M_{\odot}$ (central density:
$2.57 \times 10^{15}$gcm$^{-3}$), which is unstable.
Unlike the stable configuration --which undergoes a oscillating motion--
the unstable one collapses in a time
of the order of the characteristic dynamical time.
The behaviour of the rest-mass density, as a function of time,
can be seen in Figs. 6 and 7 for, respectively, the stable
and unstable models.
In these figures we have selected a sample of mass shells by
making a equipartition of the total gravitational mass in eleven
shells of the same width and plotted the ten inner ones.
For
the values of the central density involved in this study, the
dynamical characteristic time is of about $\le 0.8$ msec,
which is interesting in view of the different temporal scales which
govern the dynamics of both models. The stable configuration (Fig. 6)
displays a plateau profile during more than 5 times its dynamical
characteristic time. The unstable configuration (Fig. 7) starts to collapse in
a characteristic time less than two times its dynamical
characteristic time.

\vspace{1.cm}

\begin{center}
{\bf IV.4 Oppenheimer-Snyder collapse}
\end{center}

The gravitational collapse (into a black hole) of a homogeneous spherical
dust cloud (p=0) has been  exhaustively studied since the original paper of
Oppenheimer and Snyder (1939). Depending on the particular time slicing and
coordinate gauge used simple analytical expressions are available for both the
metric coefficients and the matter variables (Petrich et al., 1985, 1986). The
solution to this problem in the RGPS (as in the present work)
can be found in Petrich et al. (1986) and Gourgoulhon (1992,1993). We have
summarized these previous theoretical works in an appendix.
Since during the last epoch of Oppenheimer-Snyder collapse,
the variables involved (both the geometrical and hydrodynamical) reach
extreme values, it is generally considered a good test-bed calculation for
fully
general-relativistic time-dependent numerical codes (Petrich et al., 1985,
1986; Gourgoulhon, 1992,1993; Baumgarte et al., 1994). Hence, we have used
it as a check of our code in the case of strong gravitational fields, as well
as
ultrarelativistic velocities. Although no shocks appear during the evolution
of the presureless ball, very steep spatial and temporal gradients  develop
at the last stages of its evolution.

Results of our simulations
have been plotted together
with the exact solution in Figs. 8-10, as a function of radial coordinate,
normalized to the Schwarzschild radius of the initial configuration, for a
sample of values of the temporal coordinate labelled from $0$ to $6$ (see the
corresponding captions). Figure 8 shows the rest-mass density, normalized to
its initial value. Due to the gauge used, the existence of a
positive spatial gradient of rest-mass density is noticeable. Figure 9
displays the behaviour
of the physical velocity. As expected, the
velocity of the surface tends to unity as the collapse proceeds to its late
stage, having a minimum value of about $v=-0.92$ (plotted in Fig. 9).
The geometrical quantity
$\alpha$ is plotted in Fig. 10. At the late epoch we have succeeded in reaching
a value as low as $1.3232 \times 10^{-10}$.

In this application the position of the surface $R(t)$ is defined by
the analytical solution. Note that due to our procedure for moving
the surface the value corresponding to this point (the last numerical cell)
suffers of an error. This error is associated to the fact that our grid
is Eulerian making difficult to precise the exact
position of the surface.
A comoving grid would allow to solve the surface with more accuracy and
would result in spatial gradients steeper than
those displayed in Figs. 8-10.

Figures 11 show in a space-time diagram the trajectories of the
different mass shells in terms of the proper time (Fig. 11a) and the
coordinate time (Fig. 11b). The effect of freezing in the evolution of the
system is
noticeable from Fig. 11a, by comparison with Fig. 11b.
Our calculation has evolved to an epoch later than that reached by
Gourgoulhon (1992) with the pseudospectral technique.
The results displayed in
Figs. 8-11 confirm the powerful capabilities of our numerical techniques for
treating steep spatial and temporal gradients.

\begin{center}
{\bf IV.5 An application to the stellar core collapse}
\end{center}

A very simple way of modelling the essential features of
 stellar core collapse in massive stars is to incorporate
a simple equation of state into a hydro-code, like that
of an ideal gas, but taking an adiabatic exponent which can depend on
density according to a particular prescription.
The EOS we have used is a $\Gamma$-law such that $\Gamma$ varies with
density according to:
\begin{equation}
{\Gamma} = {\Gamma_{min}} + \eta({\log{\rho}-\log{\rho}_{b}}) \,\,\,\,,
\end{equation}
\noindent
with: $\eta=0$ if $\rho < \rho_{b}$ and $\eta>0$ otherwise (Van Riper, 1979).

Two set of values for the parameters $\Gamma_{min}$, $\eta$ and $\rho_{b}$ have
been considered: \{1.33, 1, 2.5$\times 10^{14}$ gcm$^{-3}$ \} ({\it model A})
and \{1.33, 5, 2.5$\times 10^{15}$ gcm$^{-3}$ \} ({\it model B}). Model A
exhibits standard values of the parameters, i.e., the effective adiabatic
exponent of infalling material and the value of nuclear density (for
symmetric nuclear matter at zero temperature) at the saturation point. Model
B is rather exotic due to the particular values for the bounce density and
stiffness $\eta=5$; however, we have considered
this model in order to check the ability of our hydro-code
in solving numerically flows which develop strong shocks in very
strong gravitational fields.

The initial model in the present application is a white dwarf having a
gravitational mass of $1.3862 M_{\odot}$ corresponding to the maximum
mass cited in the section before.

We have run models A and B with a grid of 330 points distributed -- as in the
previous application -- in a geometric progression
(in the following, figures are numbered with a number followed by $a$ or $b$
corresponding, respectively, to cases A and B).
The solution converges quickly by taking different grids. No relevant
differences were found in the numerical results when the number of points
is increased from 330 points to 630 grid points.
Poor grids lead to results which differ from
the converged ones, but the essential features of the shock remain
sharply solved, in typically two numerical cells. This property is one of
the most
relevant of these numerical techniques.
As in Section $\S IV.3$ the value of $R(t)$
is obtained from the condition: $\rho (R(t)) = \rho (R(t=0))$.

Table 3 shows the main features of both models during the {\it infall epoch}.
Important numbers are those corresponding to the maximum of velocity, that is,
0.41 and 0.62, respectively, for cases A and B (see also Figs. 12). Also, it
is interesting to point out the particular values of the geometric factor
$\alpha^2$ at the maximum compression and to compare them with those
corresponding at the surface of a typical neutron star $\approx 0.75$.

The kinetic energy of the material ejected has been calculated at different
times, those corresponding to some fixed values for the position of the shock
in our Eulerian frame. Table 4 displays the main features of both models during
the {\it prompt phase}. Case B is globally more energetic than case A.
The ulterior evolution after the shock has arrived at the surface has
not been followed up in the present work.

Note the behaviour of the velocity field as it can be seen in
Figs. 12. The shock is
sharply solved in one or two zones and is free of spurious oscillations. In
these figures, labels stand for the temporal sequence of each curve (see the
corresponding table caption). From Figs. 12 we can see that the
radius of the inner core, at the time of maximum compression and for which
the infall velocity is maximum (see Table 3) and the shock has been formed, is
12.6 km. At the corresponding time in case B, the size is only 6.3 km. This
strong difference together with other elements such as the particular values of
the velocity and the low values of the geometric factor $\alpha^2$ (see Table
3) are the signatures of the fact that general relativistic
effects are more important in case B than in case A.

Rest-mass density, as a function of radial coordinate, for several times is
plotted in Figs. 13. The propagation of the shock can be followed in this
figure.
Qualitatively, case B displays the same behaviour. The jump in density is of
about one order of magnitude. The maximum values of the central density differ
in both cases by a factor of five (see Table 3).

Figures 14 show several snapshots of the internal specific energy (in units of
$c^2$) as a function of radial coordinate. The shock is sharply solved and its
propagation is well defined there. Note the huge values of
the internal energy reached at the center (curve labelled by 3 in Figs. 14)
of $\approx 0.18$ (case A), being $\approx 0.42$ in case B.

The conservative features of our hydro-code, consistent with the conservation
laws of baryonic mass and gravitational mass (or binding energy), are
displayed in Figs. 15 and 16. Figures 15 show several snapshots
of the gravitational mass, as a function of radial coordinate.
The binding energy has been plotted, as a function of radial coordinate, in
Figs. 16. In both figures, curves labelled by 1 correspond to the initial
model, and
the ones labelled by 3 to the time of maximum compression. A glance at these
figures gives confidence in the conservative features of the code, since the
relative errors at the surface are consistent with the accuracy of our
algorithm. Let us focus on those curves labelled by 3: Case A generates an
inner core (the seed of the protoneutron star) which has a mass and radius
greater than those corresponding to case B.
The size in mass of the inner core at the time of maximum
compression is $\approx 1.15 M_{\odot}$ and $\approx 1.0 M_{\odot}$ for cases A
and B, respectively. The total binding energy of the initial model is, in units
of $M_{\odot} c^2$, $-3.1 \times 10^{-4}$. Conservation of binding energy is
preserved by our code.

We have built up spacetime diagrams in order to simplify the understanding
of the evolution and, eventually, to compare with other calculations. Figs. 17
show the areal radii which enclose a sample of mass shells (see captions of
these figures) in terms of the time coordinate.
Unlike the Oppenheimer-Snyder case (Fig. 11a), in Figs. 17 is very
remarkable the existence of an absolute minimum of the radius enclosing
a given mass, as well as the propagation of the shock. From these figures
we can distinguish between the inner core which reaches hydrostatic
equilibrium and the outer part which is ejected with a kinetic
energy  greater than $4 \times 10^{51}$ erg
($7 \times 10^{51}$ erg) in case A (B) when the shock is at 560 km
(see table 4).

Finally, to emphasize
the importance of general relativistic
effects we plot (in Figs. 18) the geometrical quantity
$\alpha^2$ as a function of radial coordinate and at several times of its
evolution. As can be seen from Figs. 18, $\alpha^2$ is a continuous function
throughout the shock. The curve labelled 2 corresponds to
the time at which the absolute minimum is reached at the center, being
0.49 and 0.14 for cases A and B, respectively.

\begin{center}
{\bf VII. CONCLUSIONS}
\end{center}

In this work we have described a full general relativistic one-dimensional
hydrodynamical code which incorporates a modern high-resolution
shock-capturing
algorithm for the correct modelling of formation and propagation of strong
shocks. Strong shocks are sharply solved.
Our algorithm is conservative, monotone and upwind. It makes use of
a linearized Riemann solver.
The present version of our
hydro-code has the fundamental property of
conservation of those quantities (such as the baryonic mass and the total
energy) whose evolution is described by continuity-like equations.

We have carried out several numerical tests and
applications of our code.
They have been selected in order to check a large range
of different properties:
special relativistic effects, geometrical effects, general
relativistic effects in a background or fully general relativistic dynamics.
In particular, the spherical shock
reflection problem has been
solved in the ultrarelativistic regime most successfully.
We have reproduced some of the stationary solutions describing a
flow evolving in a given background. We have compared dynamical study
of perturbations with the static stability criterion of equilibrium
configurations. Oppenheimer-Snyder collapse has given the
opportunity to check our code with the analytical solution of a
fully dynamical spacetime. Finally, we have made a simple application
to the dynamics of the collapse of compact objects using a simple
microphysics.
Case B of this last application displays strong
shocks evolving in presence of gravitational fields so huge
that the coupling introduced by the geometrical quantities
makes the system to be solved a highly nonlinear system. Hence,
the
severity of some of the tests that our code has overcome lead us to be
confident in the quality of the results in future applications.

Let us give some examples
of the astrophysical applications that we are envisaging.
First, we are interested in carrying out
simulations of collapsing stellar cores with a realistic
initial model and an updated microphysics. The influence of the
gravitational field in the formation and propagation of relativistic
fireballs -- considered as good candidates for modelling $\gamma$-ray
bursts -- will also be studied with the hydro-code analyzed
in present paper.

\begin{center}
{\bf ACKNOWLEDGMENTS}
\end{center}

This work has been supported by the  Spanish DGICYT (grant PB91-0648)
and a grant from the IVEI. Jos\'e M$^{\underline{\mbox{a}}}$.
Mart\'{\i} has benefited from a european postdoctoral fellowship (contract
number ERBCHBICT930496). Calculations were carried out in a VAX 6000/410 at the
Instituto de F\'{\i}sica Corpuscular and in a IBM 30-9021 VF at the Centre
d'Inform\`atica de la Universitat de Val\`encia. Authors acknowledge
to an anonymous referee for his useful comments and suggestions and
to Dr. Carinhas for his careful reading of the manuscript.

\vspace{1.cm}

\begin{center}
{\bf Appendix A. Oppenheimer-Snyder collapse in RGPS coordinates}
\end{center}

The gravitational collapse of a homogeneous and presureless ball of dust,
initially at rest (Oppenheimer and Snyder, 1939), has been proposed as a
test-bed for fully general-relativistic codes. Let us summarize in this
appendix the analytical results of this problem when they are expressed
in terms of the RGPS coordinates (Petrich et al. 1986, Gourgoulhon 1993).

According to Petrich et al. (1986), the solution consists of finding
coordinate transformations from the canonical forms of the
interior Friedmann --closed-- and the exterior Schwarzschild metrics.

The closed Friedmann metric is
\begin{equation}
ds^{2} =  - d \tau^{2} + a(\tau)^2 \left(d \chi^2 + \sin^2 \chi
(d \theta^2 + \sin^2 \theta d\phi^2) \right) \,\,,
\label{RW}
\end{equation}
\noindent
where $\tau$ is the proper time of the fluid and $\chi$ varies into
the interval $0 \le \chi \le \chi_s$, $\chi =0$ and $\chi_s$ being,
respectively, the values of $\chi$ at the center and at the surface of the
star. $\chi_s$ can be determined from the gravitational mass of the cloud
$M$ and its initial radius $R(0)$:
\begin{equation}
\chi_s =  \arcsin \left({\displaystyle{\frac{2M}{R(0)}}}  \right)^{1/2} \,\,.
\label{chis}
\end{equation}

In terms of the conformal time $\eta$, defined by the relation
\begin{equation}
\frac{d\eta}{d\tau} =  \frac{1}{a(\tau)} \,\,,
\end{equation}
\noindent
the metric (\ref{RW}) can be written
\begin{equation}
ds^{2} = a(\eta)^2 \left( - d \eta^{2} +  d \chi^2 + \sin^2 \chi
(d \theta^2 + \sin^2 \theta d\phi^2) \right) \,\,.
\label{RW1}
\end{equation}

The exterior metric is given by the well-known Schwarzschild solution
for vacuum
\begin{equation}
ds^{2} =  - \left( 1 - \frac{2M}{r} \right) dt^{2} +
\left( 1 - \frac{2M}{r} \right)^{-1} dr^{2} +
r^2 (d \theta^2 + \sin^2 \theta d\phi^2) \,\,.
\label{Sch}
\end{equation}

The RGPS form of the interior metric is given by Eq. (\ref{RGPS}).

The solution of the Oppenheimer-Snyder in the RGPS spacetime can be
derived in, basically, two steps: i) First, the interior Friedmann metric
(\ref{RW1}) has to be transformed into RGPS coordinates (\ref{RGPS}).
ii) Second, interior and exterior solutions have to be appropriately matched
at the surface of the star.

The explicit expressions we were looking for are:

1) {\it Coordinate transformations}:

\begin{equation}
t = 2M \left[{\displaystyle{\frac{1}{\tan \chi_s} }} \left(
\eta_s + \pi +  {\displaystyle{\frac{1}{2 \sin^2 \chi_s} }}
(\eta_s + \pi - \sin \eta_s) \right) +
\ln \left({\displaystyle{\frac{\tan (\eta_s/2) - \tan \chi_s}
{\tan (\eta_s/2) + \tan \chi_s} }} \right) \right] \,\,,
\end{equation}
\noindent
where $\eta_s= \eta_s(\eta_c)$ =
$-2 \arccos \left({\displaystyle{\frac{\cos(\eta_c(t)/2)}
{\cos^{1/2}\chi_s}}}\right)$, and
$\eta_c= \eta_c(t)$ being the value of $\eta$ at $\chi = 0$ on
the hypersurface $\Sigma_t$ (t=constant). The above equation allows
to obtain --by inverting it-- the function $\eta_c(t)$.

The radius at the surface satisfies:
\begin{equation}
R(t) = R(0) \left( 1 - {\displaystyle{\frac{\cos^2(\eta_c(t)/2)}{\cos \chi_s}}}
\right)
\end{equation}

The function $\chi(t,r)$ is implicitly defined by:
\begin{equation}
r = R(0) \left( 1 - {\displaystyle{\frac{\cos^2(\eta_c(t)/2)}{\cos \chi}}}
\right) \frac{\sin \chi}{\sin \chi_s}
\end{equation}

2) {\it Geometric quantities}:

\[
{\alpha(r,t)} = \left\{ \begin{array}{ll}
{\alpha_o}({\eta}_c, {\chi}_s) {\displaystyle{\frac{\cos{\chi} -
\cos^2({\eta}_c/2)}{(\cos^3{\chi}-\cos^2({\eta}_c/2))^{1/2}} }} &
\mbox{if $0 \le r \le R(t)$}  \\ \\
\left( 1 - {\displaystyle{\frac{2M}{r} }} \right)^{1/2} &
\mbox{if $r \ge R(t)$}
                        \end{array}
                  \right. \,\,\,\, \]

\[
X(r,t) = \left\{ \begin{array}{ll}
\left( {\displaystyle{\frac{\cos{\chi} - \cos^2({\eta}_c/2)}
{\cos^3{\chi}-\cos^2({\eta}_c/2)}  }} \right)^{1/2} &
\mbox{if $0 \le r \le R(t)$}  \\  \\
\left( 1 - {\displaystyle{\frac{2M}{r} }} \right)^{-1/2} &
\mbox{if $r \ge R(t)$}
                        \end{array}
                  \right. \,\,\,\, \]

\noindent
where
\begin{equation}
\alpha_o(\eta_c, \chi_s) \equiv {\displaystyle{\frac
{\cos^3\chi_s-\cos^2(\eta_c/2)}{(\cos\chi_s - \cos^2({\eta}_c/2))^{3/2}} }}
\end{equation}

3) {\it Matter variables}:

\begin{equation}
v(r,t) = - {\displaystyle{\frac
{ \cos(\eta_c/2) \tan \chi}{(\cos\chi - \cos^2({\eta}_c/2))^{1/2}} }}
\end{equation}

\begin{equation}
\rho(r,t) = {\displaystyle{
\frac{3 \sin^6 \chi_s}{32 \pi M^2} \left(
\frac { \cos\chi}{\cos\chi - \cos^2({\eta}_c/2)} \right)^3 }}
\end{equation}

\newpage

\begin{center}
    \begin{tabular}{|c|cccc|}
    \multicolumn{5}{c}{\bf Table 1 \rm} \\
    \multicolumn{5}{c}{\bf Relativistic spherical shock reflection \rm} \\
    \multicolumn{5}{c}{}\\
    \hline \hline
              $v_0$&$W_0$ &$\rho^+ $ & $\varepsilon^{\rm max}_r$&
$\bar{\varepsilon}_r$\\
    \hline
            $0.1$       & $1.005$  & $342$    & $0.088$ & $0.016$\\
    \hline
            $0.9$       & $2.3$    & $343$    & $0.090$ & $0.015$\\
    \hline
            $0.99$      & $7.1$    & $614$    & $0.11$ & $0.018$\\
    \hline
            $0.999$     & $22$     & $1580$   & $0.13$ & $0.020$\\
    \hline
            $0.9999$    & $71$     & $4671$   & $0.14$ & $0.021$\\
    \hline
            $0.99999$   & $224$    & $14455$  & $0.14$ & $0.021$\\
    \hline
            $0.999999$  & $707$    & $45399$  & $0.14$ & $0.022$\\
    \hline
            $0.9999999$ & $2236$   & $143252$ & $0.14$ & $0.022$\\
\hline \hline
\multicolumn{5}{c}{}\\
\multicolumn{5}{c}{}\\
\end{tabular}

Initial inflow velocity is $-v_0$ (in units of the speed of light)
corresponding to an initial Lorentz factor $W_0$.
Maximum and mean relative
errors of the postshock density,
$\varepsilon^{\rm max}_r$ and $\bar{\varepsilon}_r$ (after 1000 timesteps).

\vfill
\end{center}

\vspace{1.cm}

\begin{center}
    \begin{tabular}{|c|c|cc|cc|cc|}
    \multicolumn{8}{c}{\bf Table 2 \rm} \\
    \multicolumn{8}{c}{\bf Spherical accretion onto a black hole \rm} \\
    \multicolumn{8}{c}{}\\
    \hline \hline
            &      &    D              &
                   &    S              &
                   & $\tau$            &  \\
\hline
       $EOS$&$Grid$&$\varepsilon^{max}$&$\bar{\varepsilon}$
                   &$\varepsilon^{max}$&$\bar{\varepsilon}$
                   &$\varepsilon^{max}$&$\bar{\varepsilon}$ \\
    \hline
       Dust & 50 & 0.020 & 0.008 & 0.020 & 0.009 & 0.022 & 0.010  \\
      (p=0)& 100 & 0.006 & 0.002 & 0.006 & 0.002 & 0.007 & 0.003  \\
   \hline
       Ideal& 50 & 0.021 & 0.009 & 0.030 & 0.010 & 0.033 & 0.012  \\
        gas& 100 & 0.006 & 0.002 & 0.009 & 0.003 & 0.010 & 0.003  \\
\hline \hline
\multicolumn{8}{c}{}\\
\multicolumn{8}{c}{}\\
\end{tabular}

Maximum and mean relative
errors of the numerical solution,
$\varepsilon^{\rm max}$ and $\bar{\varepsilon}$ (at a time 720 M).
$EOS$: equation of state.

\end{center}

\newpage

\begin{center}
    \begin{tabular}{|c|ccc|}
    \multicolumn{4}{c}{\bf Table 3 \rm} \\
    \multicolumn{4}{c}{}\\
    \multicolumn{4}{c}{\bf Features of the stellar \rm} \\
    \multicolumn{4}{c}{\bf collapse: infall epoch \rm} \\
    \hline \hline
              $Model$&$\rho^{max}_{14}$ &$-v_{max}$&$ \alpha^2_{min}$ \\
    \hline
            $ A $& 7.28 & 0.41 & 0.49 \\
    \hline
            $ B $& 34.6 & 0.62 & 0.14 \\
\hline \hline
\multicolumn{4}{c}{}\\
\multicolumn{4}{c}{}\\
\end{tabular}

$ \rho^{max}_{14}$ is the maximum of central density
at the infall epoch (in units of $10^{14}$ gcm$^{-3}$).
$-v_{max}$ is the maximum of velocity
at the infall epoch (in units of the speed of light).
$ \alpha^2_{min}$ is the minimum of the purely temporal
component of the metric tensor at the infall epoch.

\vfill
\end{center}

\vspace{2.cm}

\begin{center}
    \begin{tabular}{|c|ccc|}
    \multicolumn{4}{c}{\bf Table 4 \rm} \\
    \multicolumn{4}{c}{}\\
    \multicolumn{4}{c}{\bf Features of the stellar \rm} \\
    \multicolumn{4}{c}{\bf collapse: prompt phase \rm} \\
    \hline \hline
              $Model$&$KE_1$ & $KE_2$ & $KE_3$ \\
    \hline
            $ A $& 1.26 & 3.18 & 4.53 \\
    \hline
            $ B $& 1.18 & 4.93 & 7.11 \\
\hline \hline
\multicolumn{4}{c}{}\\
\multicolumn{4}{c}{}\\
\end{tabular}

$KE_i$ kinetic energy of the material reaching escape velocities,
when the shock is at the position $r_i$ (=100, 240, 560 km.,respectively).
Energies are given in units of $10^{51}$ erg.

\vfill
\end{center}

\newpage

{\bf Figure captions}

{\bf Figure 1.-}
Relativistic spherical shock reflection.
We use a continuous
line for the exact solution and different symbols like a plus, a diamond and a
triangle for velocity (in units of the speed of light), density and
specific internal energy, respectively.

{\bf Figure 2.-}
Spherical accretion of dust onto a Schwarzschild black hole.
Snapshots of the relativistic
mass density, in logarithmic scale and geometrized units,
as a function of the radial coordinate (in units of the mass of the
black hole)

{\bf Figure 3.-}
Spherical accretion of dust onto a Schwarzschild black hole.
Velocity versus the radial coordinate (in units of the mass of the
black hole) at a particular time.

{\bf Figure 4.-}
Velocity profile as a function of radial coordinate and time for
a stable white-dwarf.

{\bf Figure 5.-}
Velocity profile as a function of radial coordinate and time for
an unstable white-dwarf.

{\bf Figure 6.-}
Rest-mass density as a function of time for a stable model of neutron
star having $M=1.94532 M_{\odot}$ and a central density of $2.55 \times
10^{15}$ g/cm$^3$. Each curve corresponds to the following mass shells:
$m_j = (M/11)\times j$ $(j=1,...10)$

{\bf Figure 7.-}
Rest-mass density as a function of time for an unstable model of neutron
star having $M=1.94518 M_{\odot}$ and a central density of $2.57 \times
10^{15}$ g/cm$^3$. Each curve corresponds to the following mass shells:
$m_j = (M/11)\times j$ $(j=1,...10)$

{\bf Figure 8.-}
Snapshots of the rest-mass density (normalized to the initial value) as a
function of the radial coordinate (normalized to 2M) for the Oppenheimer-Snyder
collapse. Labels stand, respectively, for the following values of time (in
msec.): 0, 0.155, 0.189, 0.206, 0.218, 0.228 and 0.430

{\bf Figure 9.-}
Snapshots of the velocity as a function of the radial coordinate (normalized to
2M) for the Oppenheimer-Snyder collapse. Labels stand, respectively, for the
following values of time (in msec.): 0, 0.155, 0.189, 0.206, 0.218, 0.228 and
0.430

{\bf Figure 10.-}
Snapshots of the lapse as a function of the radial coordinate (normalized to
2M) for the Oppenheimer-Snyder collapse. Labels stand, respectively, for the
following values of time (in msec.): 0, 0.155, 0.189, 0.206, 0.218, 0.228 and
0.430

{\bf Figure 11a.-}
Spacetime diagram for the Oppenheimer-Snyder collapse.
Trajectories of a sample of mass shells in a proper time (in msec.) versus
radial coordinate (normalized to 2M).
Each curve corresponds to the following mass shells:
$m_j = (M/11)\times j$ $(j=1,...10)$

{\bf Figure 11b.-}
Spacetime diagram for the Oppenheimer-Snyder collapse.
Trajectories of a sample of mass shells in a coordinate time (in msec.) versus
radial coordinate (normalized to 2M) diagram.
Each curve corresponds to the following mass shells:
$m_j = (M/11)\times j$ $(j=1,...10)$

{\bf Figure 12a.-}
Snapshots of the velocity profile as a function of the radial coordinate (model
A). Labels stand , respectively, for the following values of time (in msec.):
80.49, 80.74, 80.91, 81.04, 81.20, 81.39, 81.76, 81.94, 82.86, 84.25 and 86.12

{\bf Figure 12b.-}
Snapshots of the velocity profile as a function of the radial coordinate (model
B). Labels stand , respectively, for the following values of time (in msec.):
80.68, 80.79, 80.93, 81.06, 81.25, 81.33, 81.76, 82.47, 83.22, 84.73 and 86.51

{\bf Figure 13a.-}
Snapshots of the rest-mass density profile as a function of the radial
coordinate (model A). Labels stand , respectively, for the following values of
time (in msec.): 80.49, 80.91, 81.20, 81.76, 82.86 and 86.12

{\bf Figure 13b.-}
Snapshots of the rest-mass density profile as a function of the radial
coordinate (model B). Labels stand , respectively, for the following values of
time (in msec.): 80.68, 80.93, 81.25, 81.76, 83.22 and 86.51

{\bf Figure 14a.-}
Snapshots of the specific internal energy profile as a function of the radial
coordinate (model A). Labels stand , respectively, for the following values of
time (in msec.): 80.49, 80.74, 80.91, 81.04, 81.20, 81.39, 81.76, 81.94, 82.86,
84.25, and 86.12

{\bf Figure 14b.-}
Snapshots of the specific internal energy profile as a function of the radial
coordinate (model B). Labels stand , respectively, for the following values of
time (in msec.): 80.68, 80.79, 80.93, 81.06, 81.25, 81.33, 81.76, 82.47, 83.22,
84.73 and 86.51

{\bf Figure 15a.-}
Snapshots of the gravitational mass profile as a function of the radial
coordinate (model A). Labels stand , respectively, for the following values of
time (in msec.): 0, 80.49, 80.91, 81.20, 81.76, 82.86 and 86.12

{\bf Figure 15b.-}
Snapshots of the gravitational mass profile as a function
of the radial coordinate (model B).
Labels stand , respectively, for the following values of time (in msec.):
0, 80.68, 80.93, 81.25, 81.76, 83.22 and 86.51

{\bf Figure 16a.-}
Snapshots of the binding energy profile as a function of the radial coordinate
(model A). Labels stand , respectively, for the following values of time (in
msec.): 0, 80.49, 80.91, 81.20, 81.76, 82.86 and 86.12

{\bf Figure 16b.-}
Snapshots of the binding energy profile as a function
of the radial coordinate (model B).
Labels stand , respectively, for the following values of time (in msec.):
0, 80.68, 80.93, 81.25, 81.76, 83.22 and 86.51

{\bf Figure 17a.-}
Spacetime diagram for model A.
Trajectories of a sample of mass shells in a proper time (in msec.) versus
radial coordinate (in km. and logarithmic scale).
Each curve corresponds to the following mass shells:
$m_j = (M/11)\times j$ $(j=1,...10)$

{\bf Figure 17b.-}
Spacetime diagram for model B.
Trajectories of a sample of mass shells in a proper time (in msec.) versus
radial coordinate (in km. and logarithmic scale).
Each curve corresponds to the following mass shells:
$m_j = (M/11)\times j$ $(j=1,...10)$

{\bf Figure 18a.-}
Snapshots of the geometrical quantity $\alpha^2$ as a function of the radial
coordinate (model A). Labels stand , respectively, for the following values of
time (in msec.): 80.49, 80.91, 81.76, and 86.12

{\bf Figure 18b.-}
Snapshots of the geometrical quantity $\alpha^2$ as a function
of the radial coordinate (model B).
Labels stand , respectively, for the following values of time (in msec.):
80.68, 80.93, 81.76 and 86.51


\newpage

\begin{thebibliography}{99}
\bibitem{A89} Anile, A.M., 1989, in  {\it Relativistic fluids and
magneto-fluids}, Cambridge University Press
\bibitem{ADM62} Arnowitt, R., Deser, S., and Misner, C.W., 1962,
in {\it Gravitation}, ed. by Witten, L., Wiley
\bibitem {BST94} Baumgarte, T.W., Shapiro, S.L., and Teukolsky, S.A., 1995,
ApJ, {\bf 443}, 717.
\bibitem {BW85}  Bethe, H., Wilson, J. R., 1985, ApJ, {\bf 295}, 14
\bibitem{BM91} Bonazzola, S., and Marck, J.A., 1991, J. Comp. Phys.,
{\bf 97}, 535
\bibitem{Bo64} Bondi, H., 1964, Proc. Royal Soc. of London, {\bf A281},
39
\bibitem {CW84} Centrella, J., and Wilson, J. R.,  ApJ Suppl. Ser., 1984,
{\bf 54}, 229
\bibitem {CW84}  Colella, P., and Woodward, P.R., J. Comp. Phys., 1984,
{\bf 54}, 174
\bibitem {CF48}  Courant, R., and Friedrichs, K.O., 1948, "{\it
Supersonic Flows and Shock Waves}, (Interscience)
\bibitem {DI851}
Diaz-Alonso, J., 1985, Phys. Rev., {\bf D 31}, 1315.
\bibitem {DIb85}
Diaz-Alonso, J., and
Ib\'a\~{n}ez, J. M$^{\underline{\mbox{a}}}$., 1985,
ApJ, {\bf 291}, 308.
\bibitem {DIH89}
Diaz-Alonso, J.,
Ib\'a\~{n}ez, J. M$^{\underline{\mbox{a}}}$., and Sivak, H.,
1989, Phys.Rev., {\bf C 39}, 671.
\bibitem {E93} Eulderink, F., 1993, Ph.D.Thesis,
Sterrewacht Leiden
\bibitem {Ev86}  Evans, C. R., in "{\it Dynamical space-times and
numerical relativity \rm}, ed. by J. Centrella (Cambridge University Press,
1986)
\bibitem{FIMM92} Font, J.A.,  Ib\'a\~nez, J.M$^{\underline{\mbox{a}}}$.,
Marquina, A., and Mart\'{\i}, J.M$^{\underline{\mbox{a}}}$., 1994,
A\&A, {\bf 282}, 304
\bibitem{GM77}  Gingold, R.A., and  Monaghan, J.J., 1977, MNRAS,
{\bf 181}, 375
\bibitem{GO77}
Gottlieb, D., and Orszag, S., 1977, {\it Numerical Analysis of
Spectral Methods.: Theory and Application}, Regional Conference Series
Lectures in Applied Mathematics,  {\bf 26} (SIAM, Philadelphia,1977)
\bibitem{Go92}
Gourgoulhon, E., 1991, A\&A, {\bf 252}, 651
\bibitem{Go92a}
Gourgoulhon, E., 1992, Ph.D. Thesis, Universit\'e de Paris VII
\bibitem {G93}
Gourgoulhon, E., 1993, Ann. Phys. Fr., {\bf 18}, 1.
\bibitem {HTWW65} Harrison, B.K., Thorne, K.S., Wakano, M., and
Wheeler, J.A., 1965,
{\it Gravitation Theory and Gravitational Collapse}, Chicago University
Press.
\bibitem{HSW84}
Hawley, J.F., Smarr, L.L.,  and  Wilson, J.R., 1984,
Ap. J. Suppl., {\bf 55}, 211
\bibitem {Hetal94}
Herant, M., Benz, W., Hix, W.R., Fryer, Ch.L.,
and Colgate, S.A., 1994, ApJ, {\bf 435}, 339
\bibitem{I84}
Ib\'a\~{n}ez, J.M$^{\underline{\mbox{a}}}$., 1984,
A\&A, {\bf 135}, 382
\bibitem{IMMR92}
Ib\'a\~{n}ez, J.M$^{\underline{\mbox{a}}}$.,
Mart\'{\i}, J.M$^{\underline{\mbox{a}}}$., Miralles, J.A., and
Romero, V., 1992, in {\it Approaches to Numerical Relativity}, ed. by
D'Inverno,
Cambridge University Press
\bibitem{I93}
Ib\'a\~{n}ez, J.M$^{\underline{\mbox{a}}}$., 1993,
in {\it Rotating Objects and Relativistic Physics}, ed. by
Chinea, F.J., and Gonz\'alez-Romero, L.M., Series Lecture Notes in
Physics, {\bf 423}, 149 (Springer-Verlag)
\bibitem{JM93}
Janka, H.-Th., and M\"{u}ller, E., 1993,
in Frontiers of Neutrino Astrophysics (Universal Academy Press: Tokyo)
\bibitem{La73}
Lax, P., 1972,  {\it Regional Conference Series Lectures in
Applied Math.},  {\bf 11} (SIAM, Philadelphia)
\bibitem{Le92}  LeVeque, R.J., 1992, in {\it Numerical Methods for
Conservation Laws}, Birkh\"auser.
\bibitem{Lu77}
Lucy, L.B., 1977, Astron. J.,{\bf 82}, 1013
\bibitem{Man91}
Mann, P.J., 1991, Comput. Phys. Comm., {\bf 67}, 245
\bibitem {MAL91}
Marquina, A.,
Mart\'{\i}, J.M$^{\underline{\mbox{a}}}$.,
Ib\'a\~{n}ez, J.M$^{\underline{\mbox{a}}}$. , Miralles, J.A.,
and Donat, R., 1992, A\&A, {\bf 258}, 566
\bibitem {Ma88}
Mart\'{\i}, J. M$^{\underline{\mbox{a}}}$., Miralles, J. A.,
Diaz-Alonso, J., and
Ib\'a\~{n}ez, J. M$^{\underline{\mbox{a}}}$., 1988, ApJ, {\bf 329}, 780.
\bibitem {MIM91} Mart\'{\i}, J.M$^{\underline{\mbox{a}}}$.,
Ib\'a\~{n}ez, J.M$^{\underline{\mbox{a}}}$. , and Miralles, J.A.,
1991, Phys. Rev., {\bf D43}, 3794
\bibitem {MM94} Mart\'{\i}, J.M$^{\underline{\mbox{a}}}$., and
M\"uller, E., 1994, J.Fluid Mech., {\bf 258}, 317
\bibitem {MM95} Mart\'{\i}, J.M$^{\underline{\mbox{a}}}$., and
M\"uller, E., 1995, J. Comp. Phys., {\it in press}.
\bibitem{MMFI95} Mart\'{\i}, J.M$^{\underline{\mbox{a}}}$.,
M\"uller, E., Font, J.A., and Ib\'a\~{n}ez, J.M$^{\underline{\mbox{a}}}$.,
1995a, ApJ Lett., {\it in press}
\bibitem{MMFIM95} Mart\'{\i}, J.M$^{\underline{\mbox{a}}}$.,
M\"uller, E., Font, J.A., Ib\'a\~{n}ez, J.M$^{\underline{\mbox{a}}}$.
and Marquina, A., 1995b {\it in preparation}
\bibitem{MW67} May, M.M., and White, R.H., 1967, Math. Comp. Phys.,
{\bf 7}, 219
\bibitem{MLR93} M\'esz\'aros, P., Laguna, P., and Rees, M.J., 1993,
ApJ, {\bf 415}, 181
\bibitem{Mi72} Michel, F.C., 1972, Ap. Space Sci., {\bf 15}, 153
\bibitem {Na81} Nakamura, T., Prog. Teor. Phys., 1981,
{\bf 65}, 1876
\bibitem {NMMS80} Nakamura, T., Maeda, K., Miyama, S., and
Sasaki, M., Prog. Teor. Phys., 1980,
{\bf 63}, 1229
\bibitem {NS82} Nakamura, T., and Sato, H., Prog. Teor. Phys., 1982,
{\bf 67}, 1396
\bibitem{N87} Noh, W.F., 1987, J. Comp. Phys., {\bf 72}, 78.
\bibitem{NW86} Norman, M.L., and Winkler, K-H.A., 1986, in
{\it Astrophysical Radiation Hydrodynamics}, ed. by
Norman, M.L., and Winkler, K-H.A. (Reidel).
\bibitem {OS39}
Oppenheimer, J.R., and Snyder, H., 1939, Phys.Rev., {\bf 56}, 455.
\bibitem {PST85}
Petrich, L.I., Shapiro, S.L., and Teukolsky, S.A., 1985,
Phys.Rev., {\bf D31}, 2459.
\bibitem {PST86}
Petrich, L.I., Shapiro, S.L., and Teukolsky, S.A., 1986,
Phys.Rev., {\bf D33}, 2100.
\bibitem {Pe89} Petrich, L.I., Shapiro, S.L., Stark, R.F., and
Teukolsky, S.A., 1989, ApJ, {\bf 336}, 313
\bibitem {Pi80}  Piran, T., J. Comp. Phys., 1980,
{\bf 35}, 254
\bibitem{P93} Piran, T., Shemi, A., and Narayan, R., 1993, MNRAS,
{\bf 263}, 861.
\bibitem {Retal91} Romero, J.V., Diaz-Alonso, J., Ib\'a\~nez,
J.M$^{\underline{\mbox{a}}}$.,
Miralles, J.A.,  and P\'erez, A., 1992, ApJ, {\bf 395}, 612.
\bibitem{Sch93} Schneider, V., Katscher, V., Rischke, D.H., Waldhauser, B.,
Marhun, J.A.,  and Munz, C.-D., 1993, J. Comput. Phys., {\bf 105}, 92.
\bibitem {ST83} Shapiro, S.L., and Teukolsky, S.A., 1983,
{\it Black Holes, White Dwarfs and Neutron Stars}, John Wiley.
\bibitem {SP87}   Stark, R. F., and Piran, T., Comp. Phys. Rept., 1987,
{\bf 5}, 221
\bibitem {Sw94}  Swesty, F.D., Lattimer, J.M., and Myra, E.S., 1994,
ApJ., {\bf 425}, 195
\bibitem{VL79} Van Leer, B., 1979, J. Comp. Phys., {\bf 32}, 101.
\bibitem{VR79}  Van Riper, K.A., 1979, ApJ, {\bf 232}, 558
\bibitem {Wi72}  Wilson, J. R., 1972, ApJ, {\bf 173}, 431
\bibitem {Wi79} Wilson, J. R., in "{\it Sources of gravitational
radiation \rm}, ed. L. L. Smarr (Cambridge University Press, 1979)
\end{thebibliography}
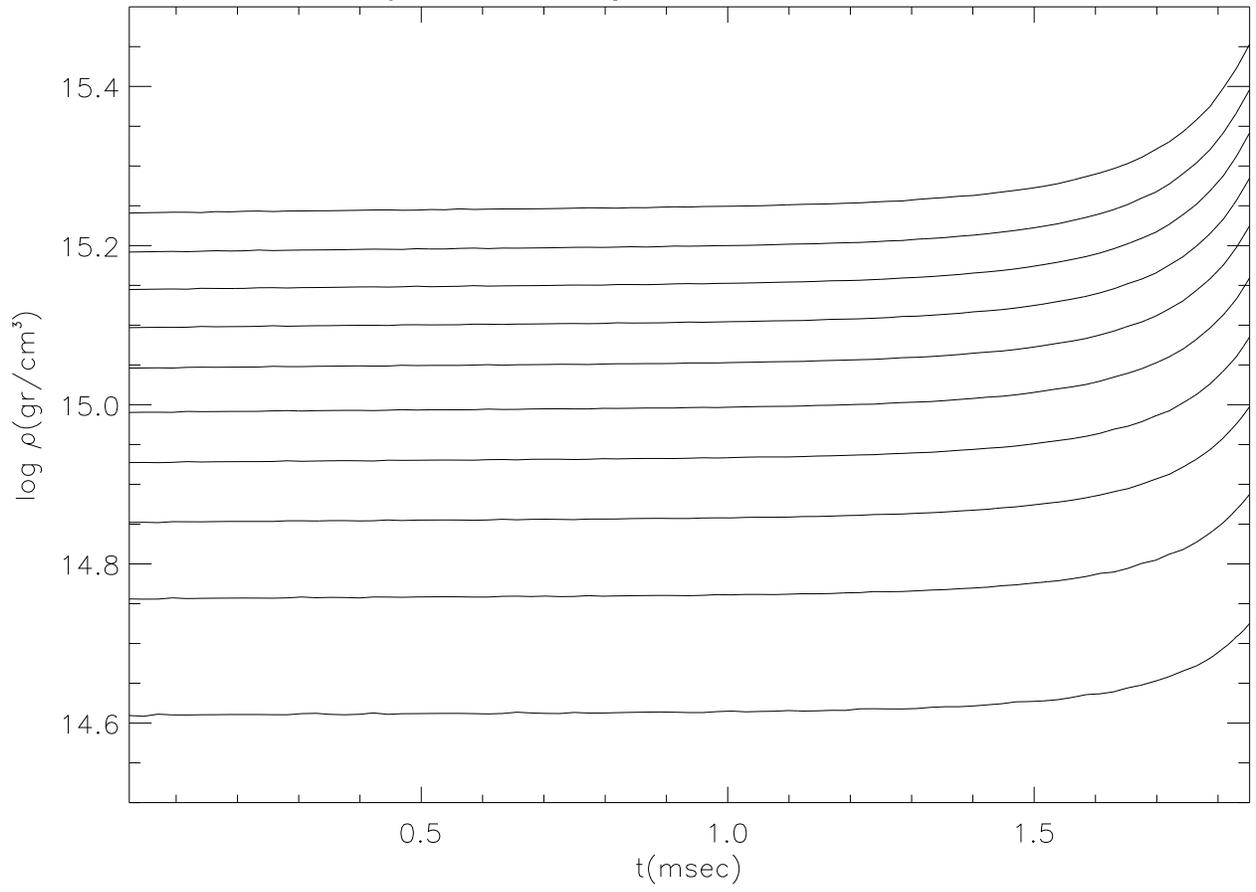
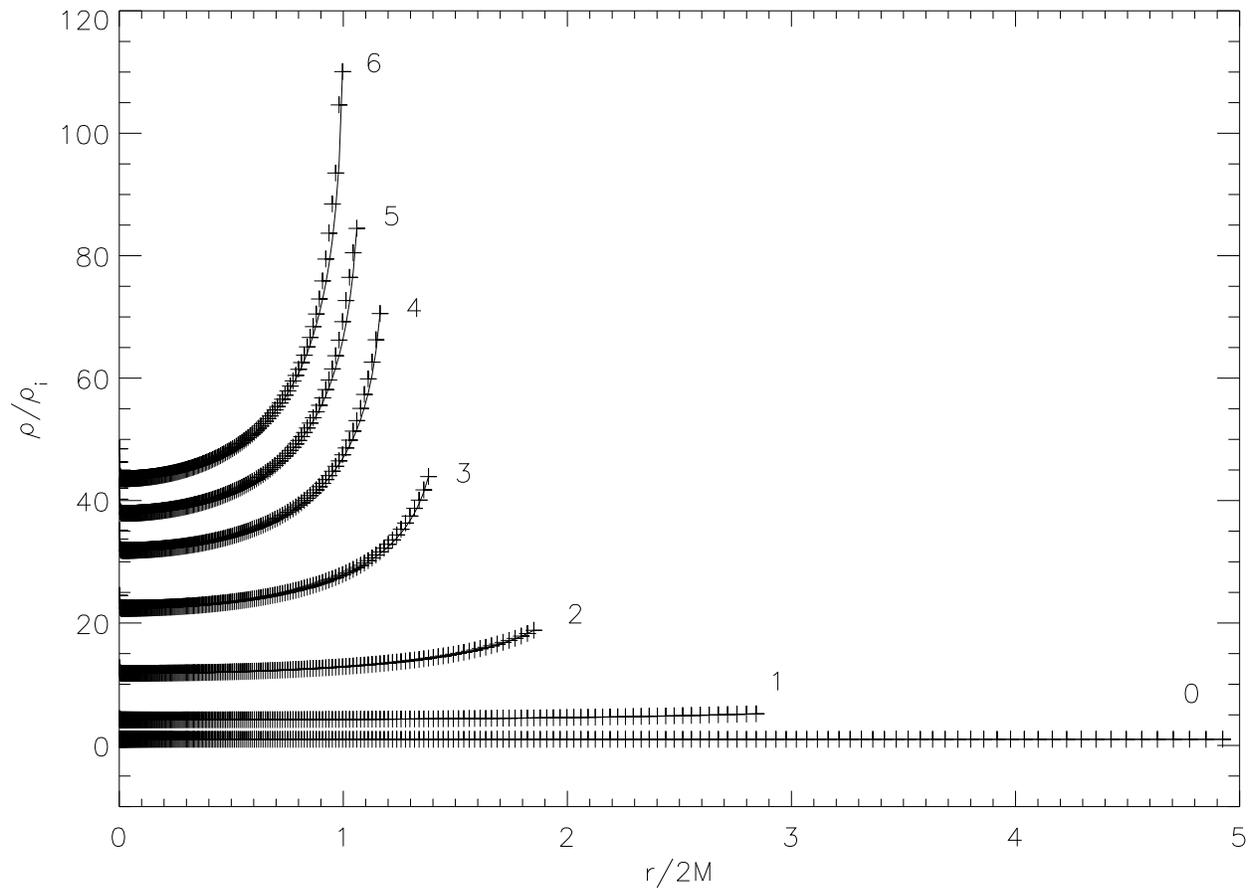
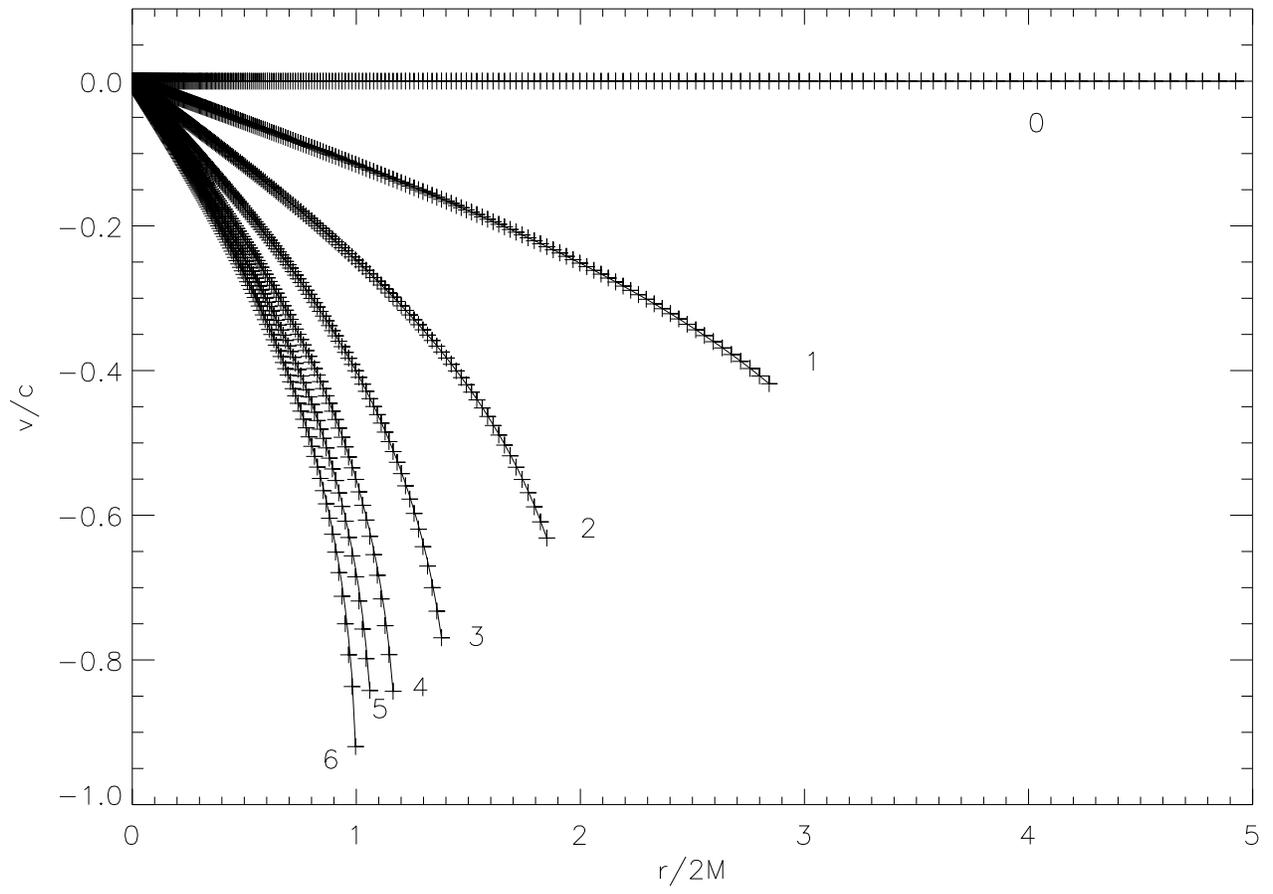
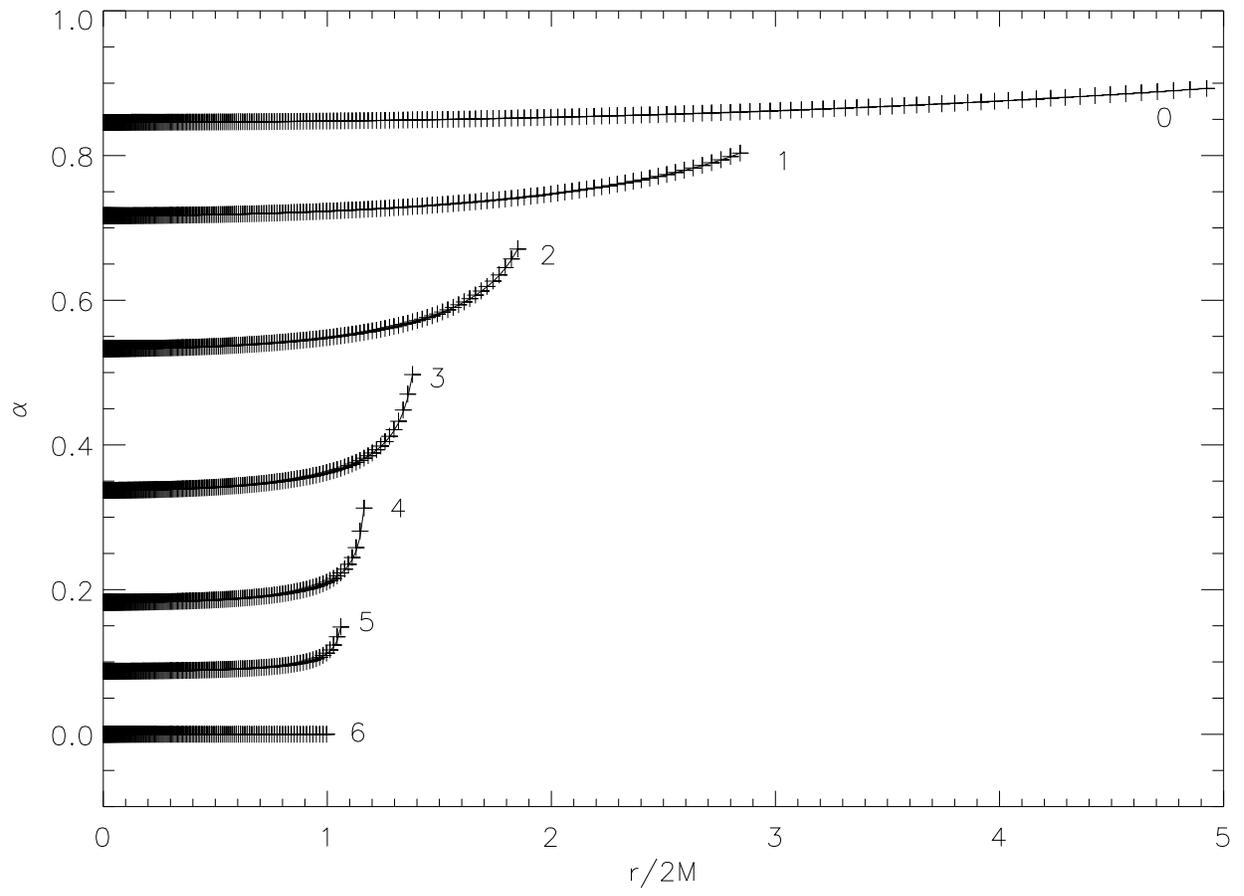
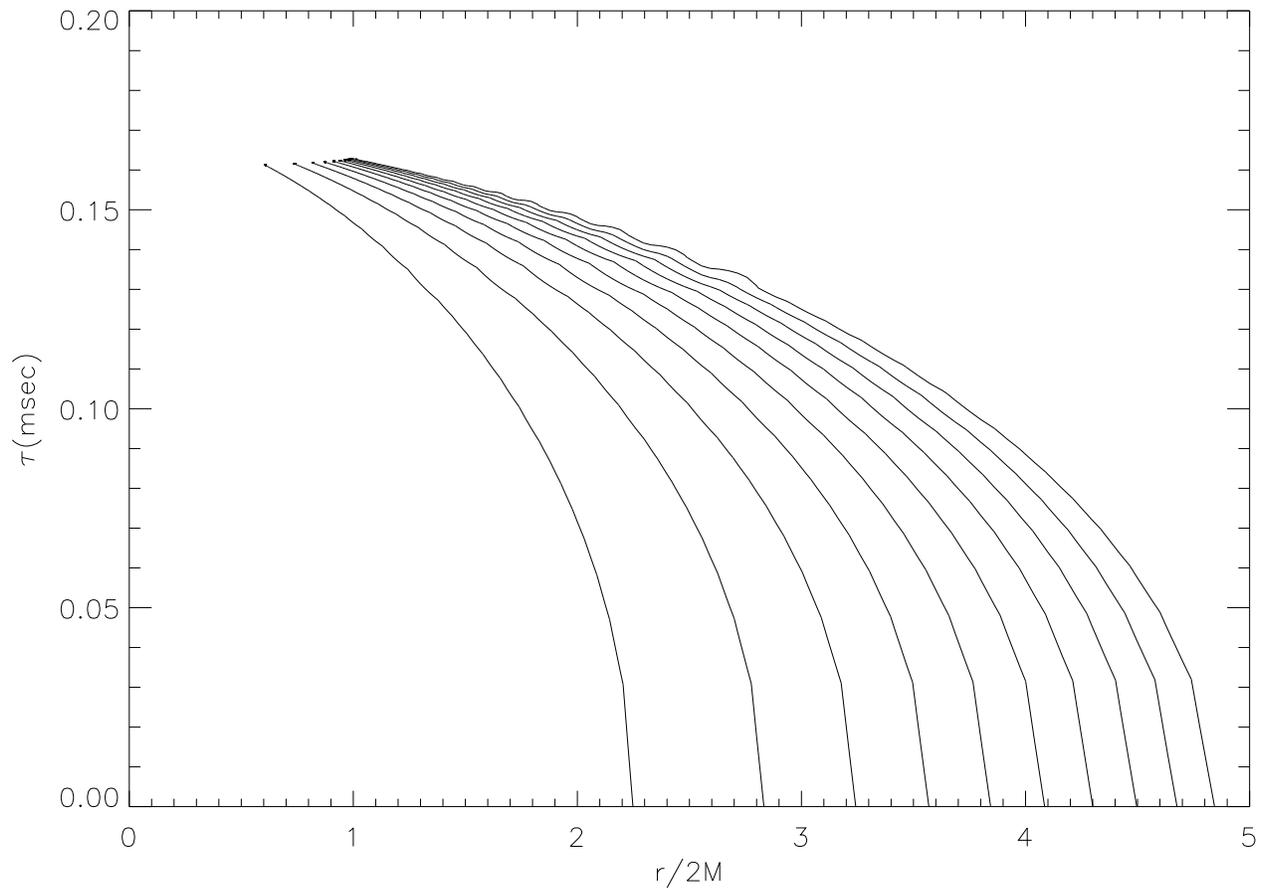
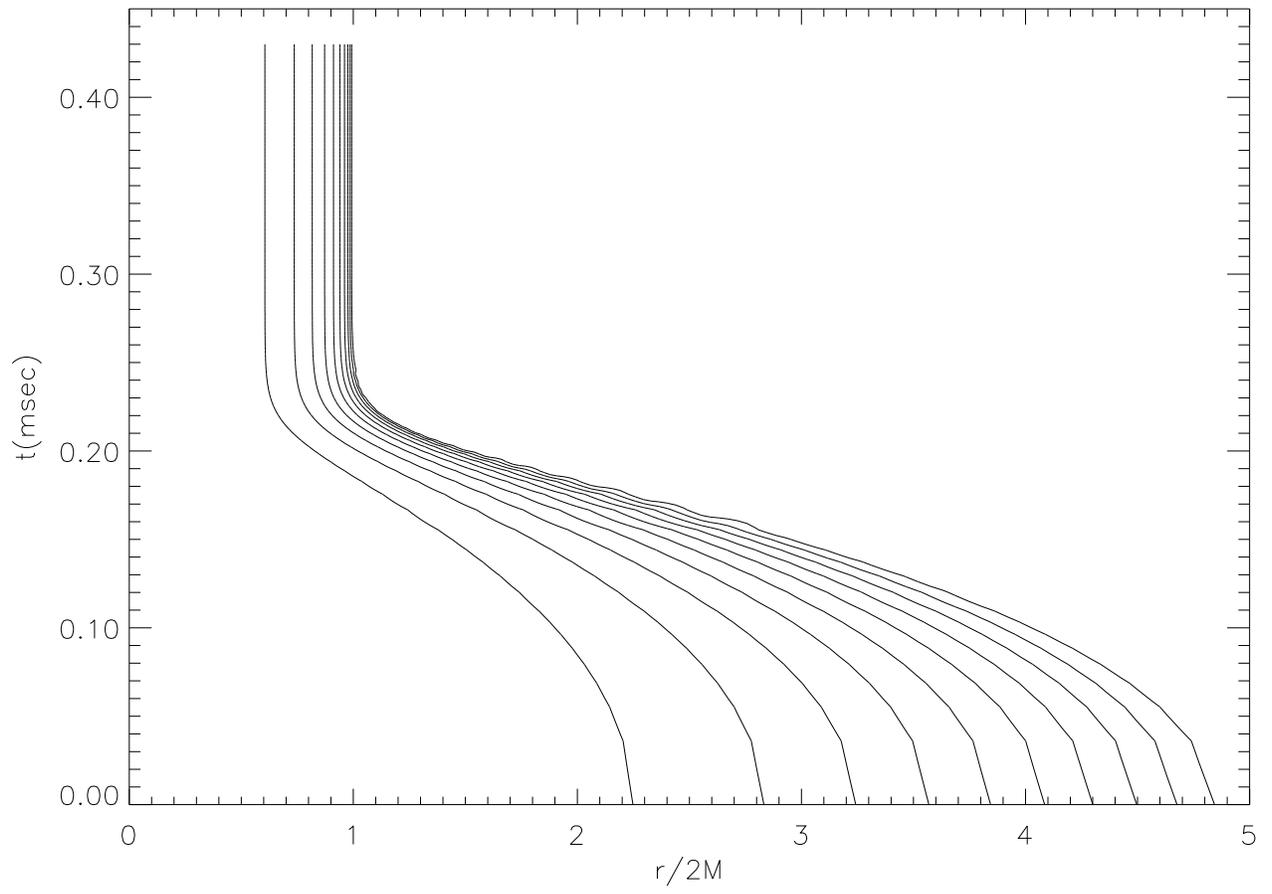
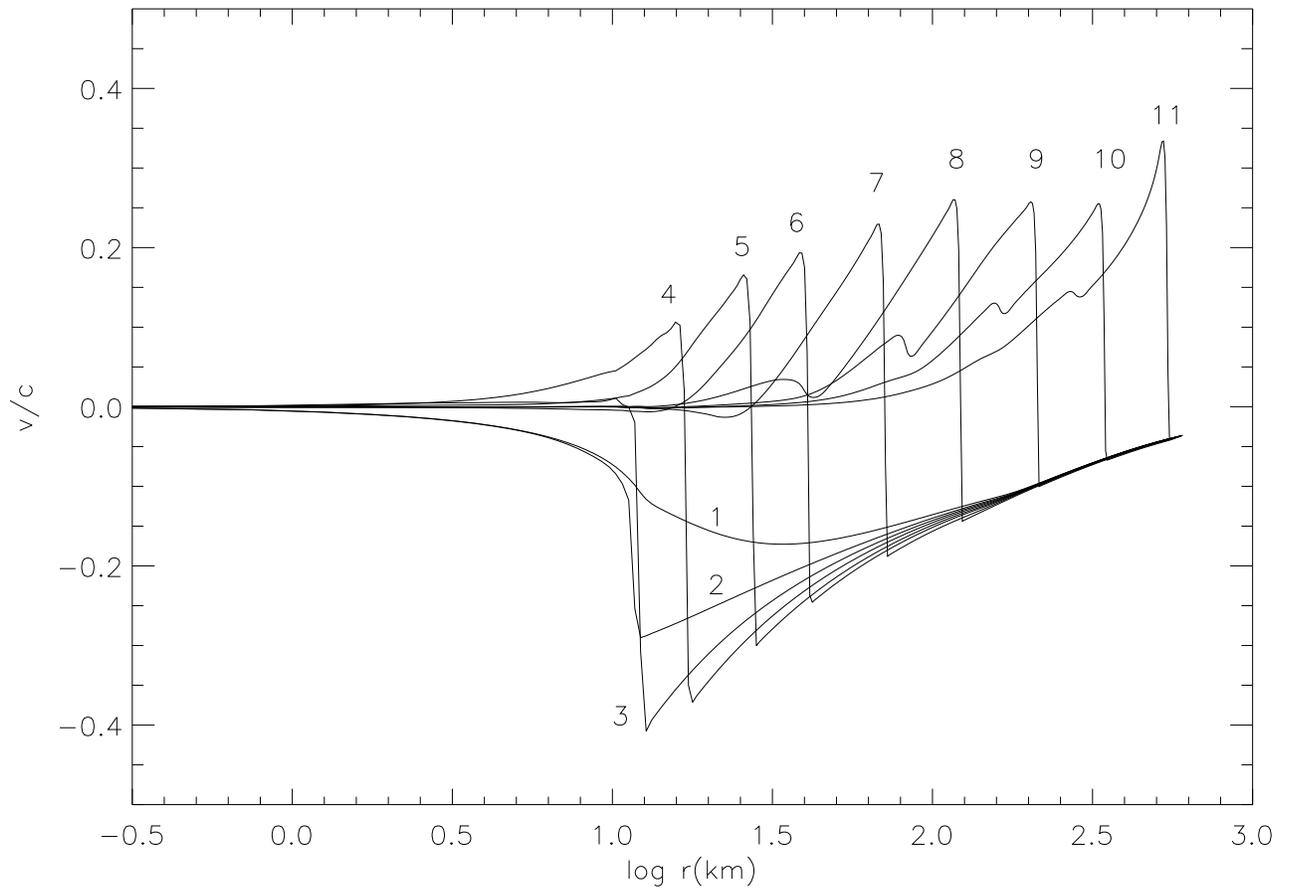
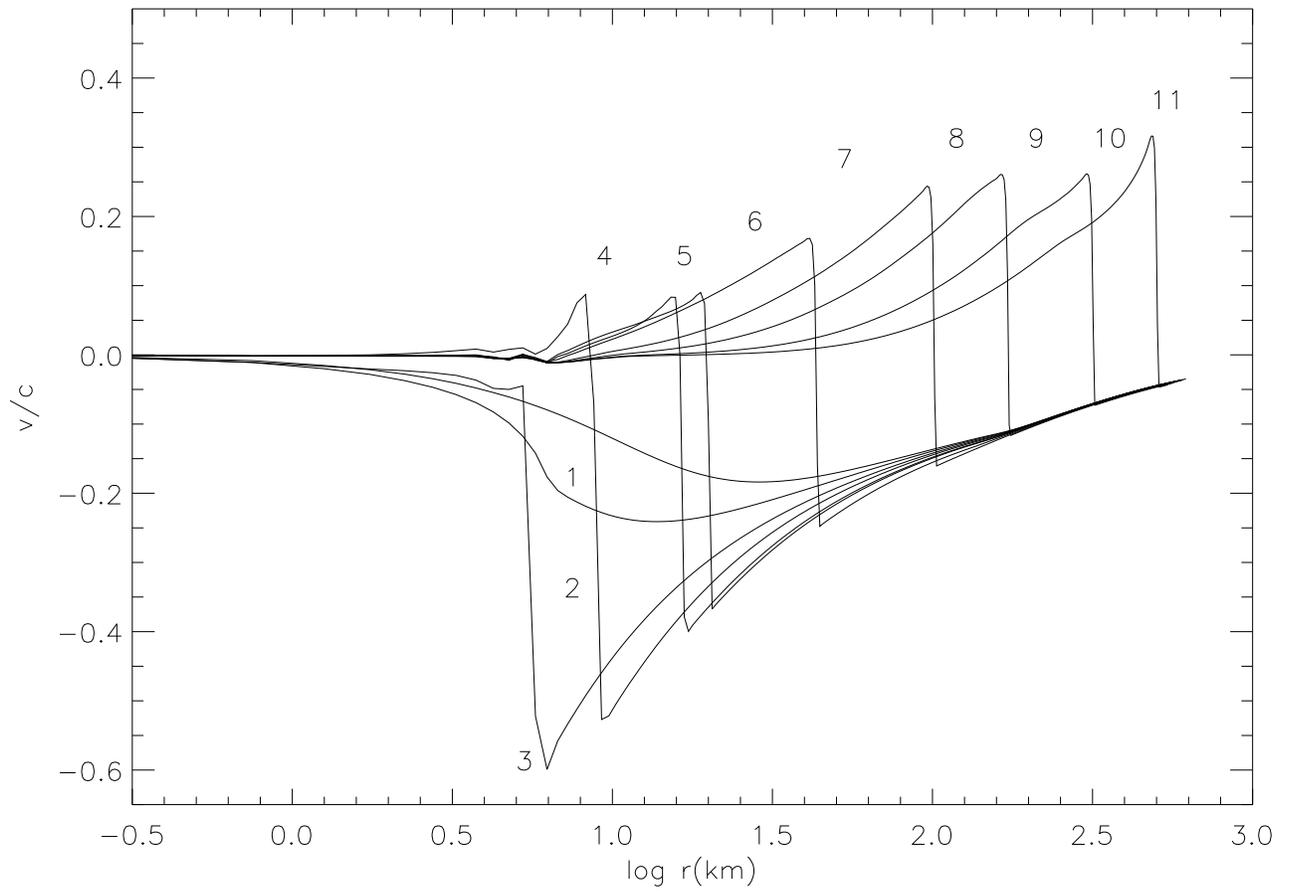
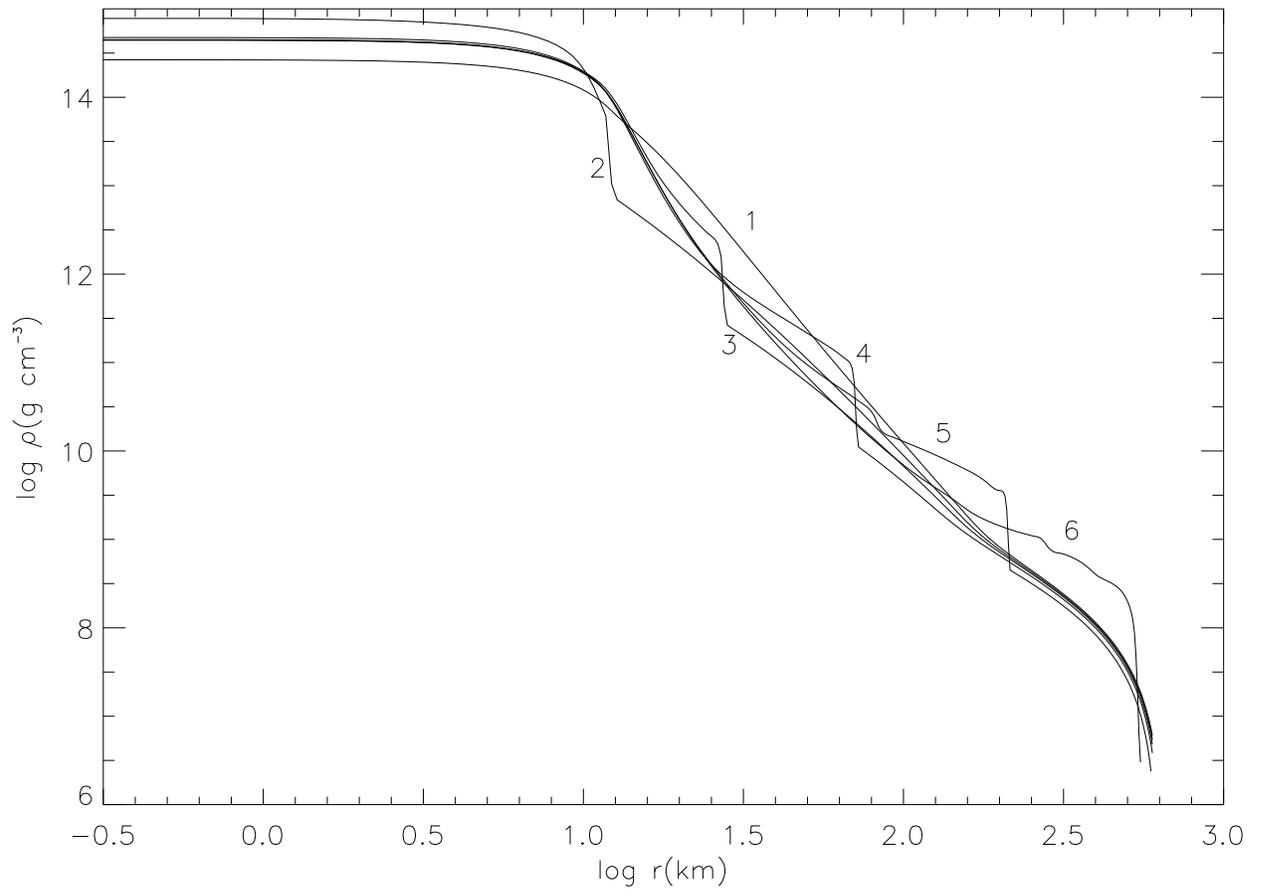
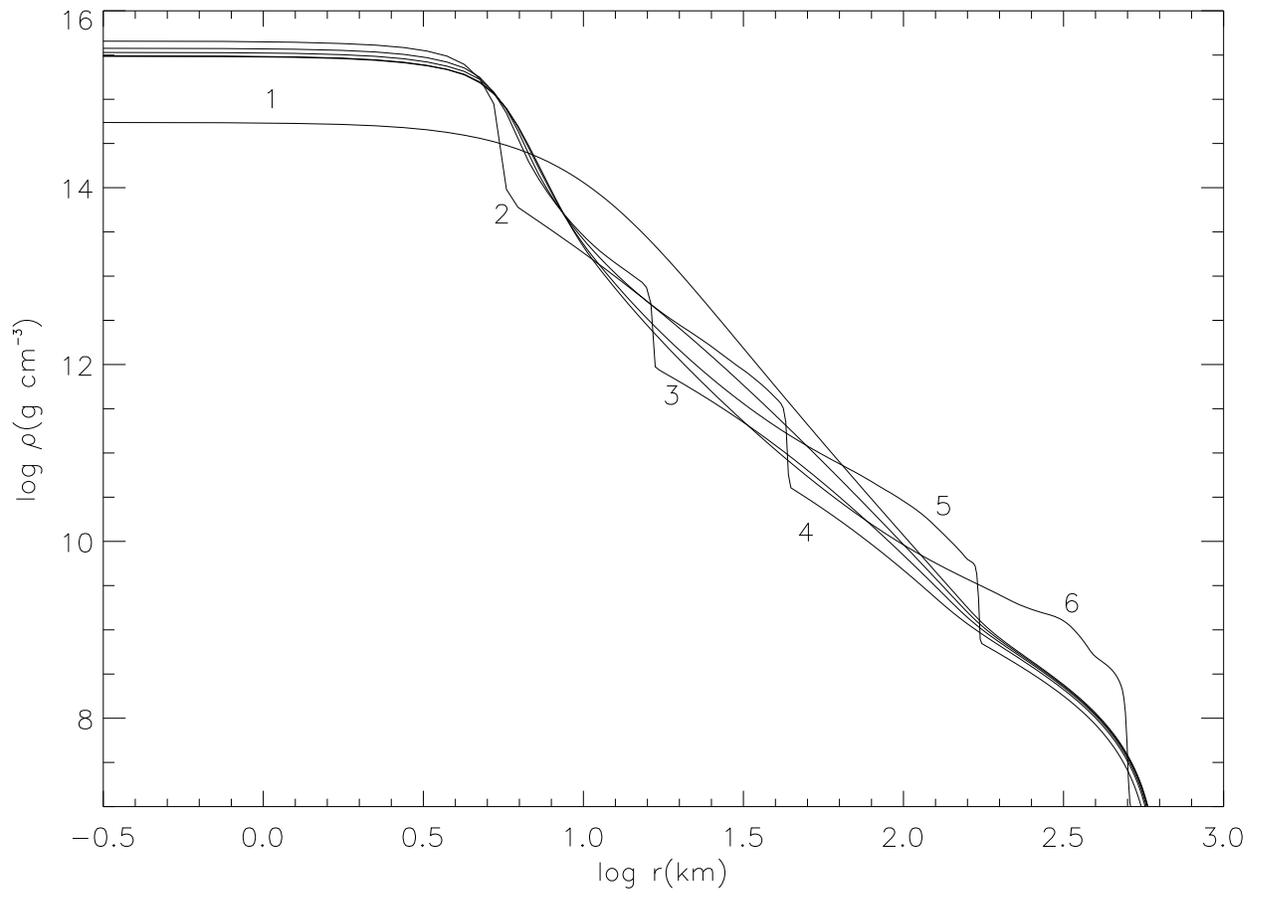
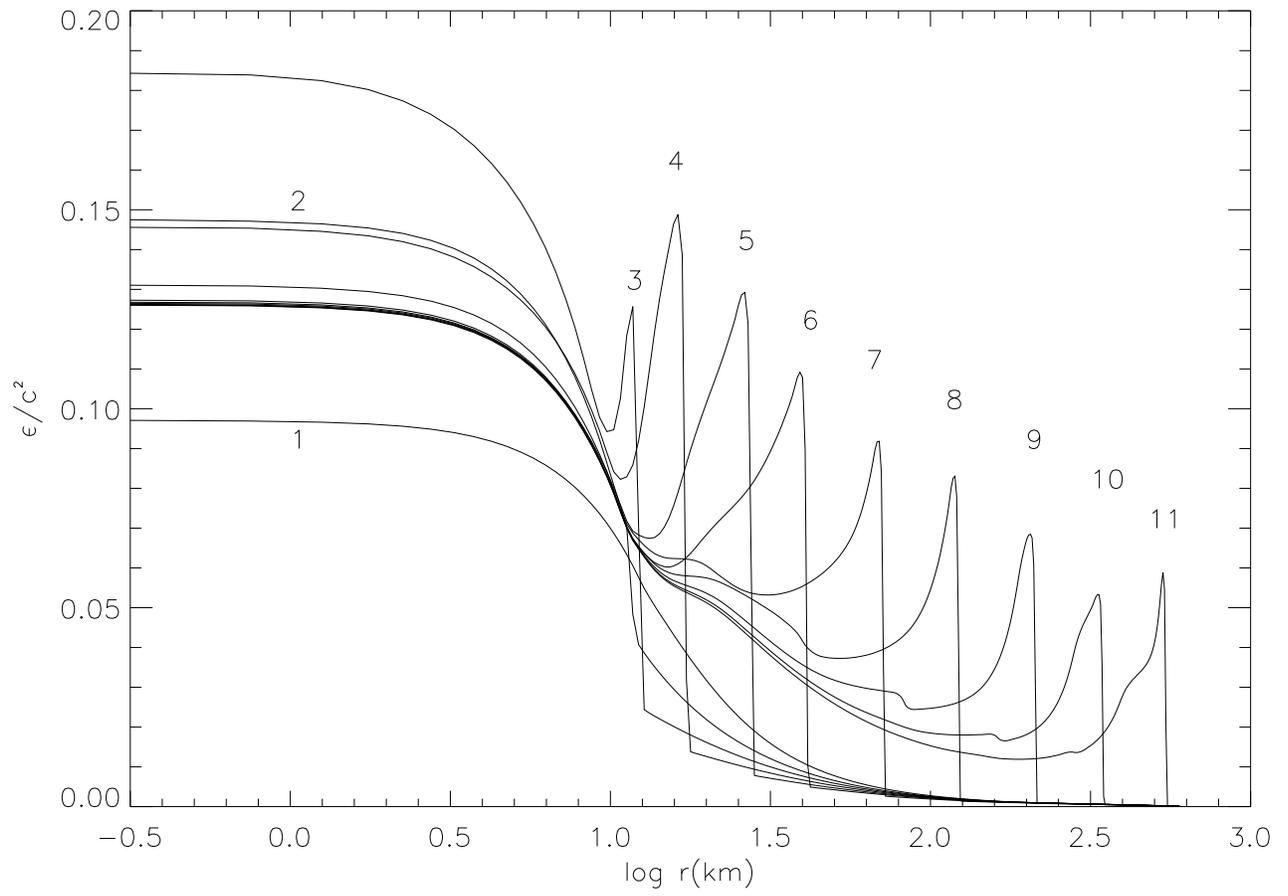
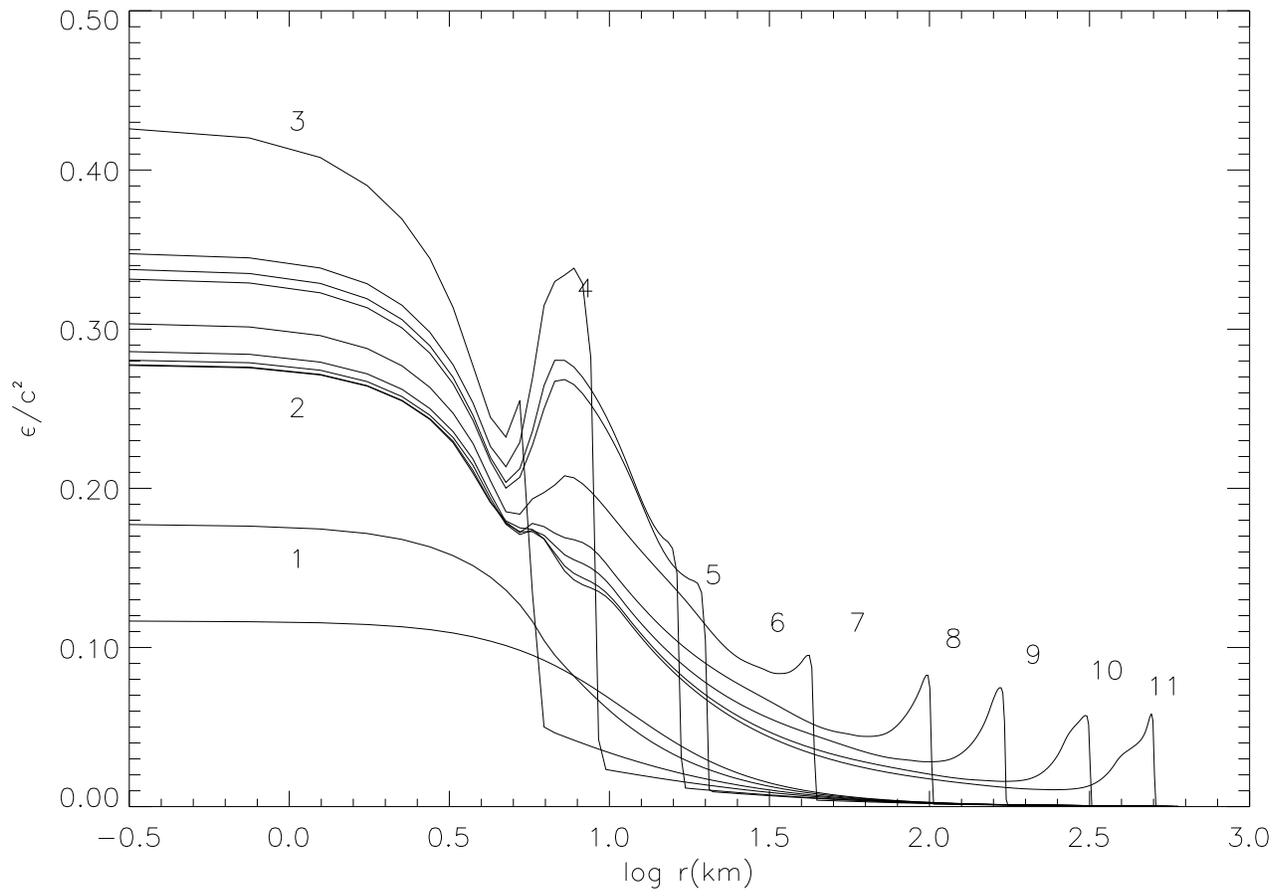
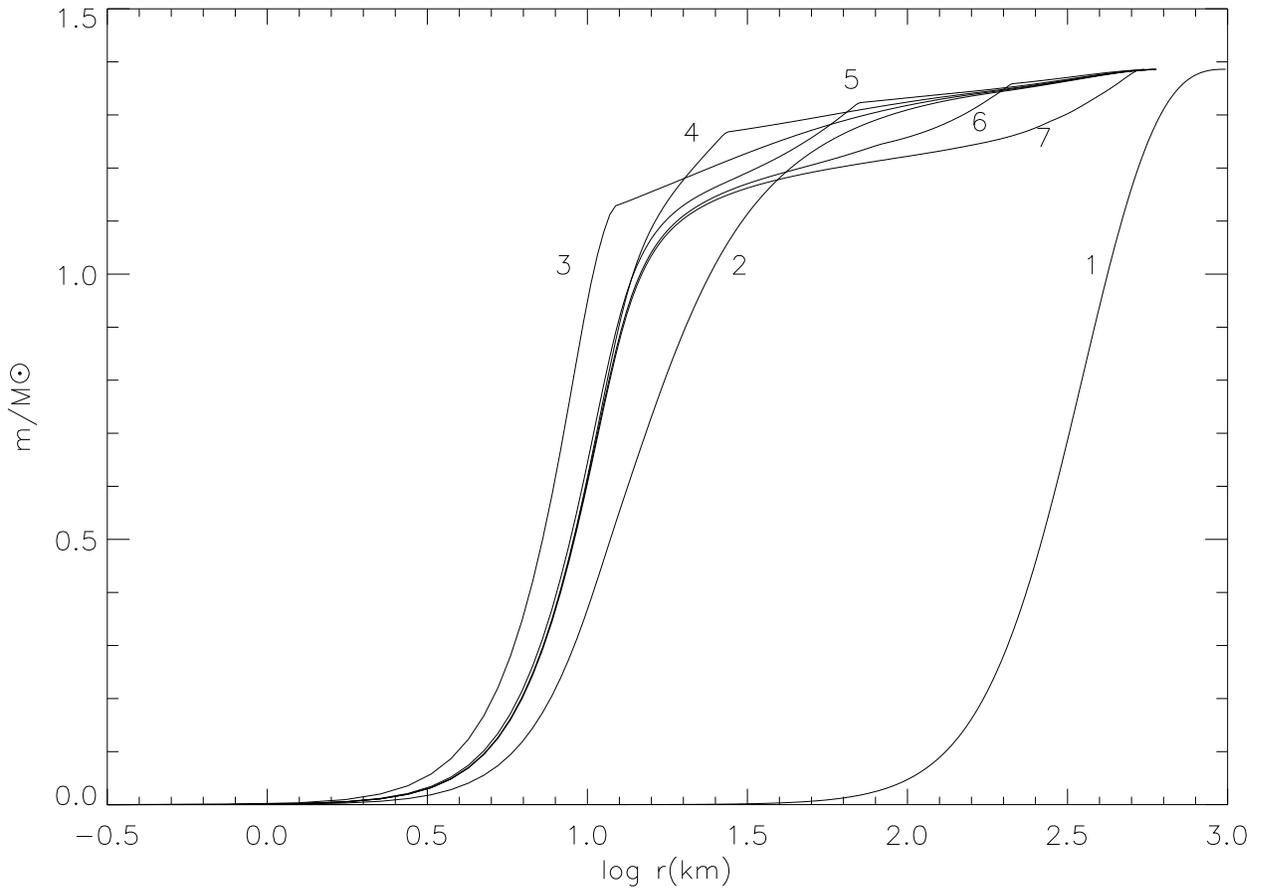
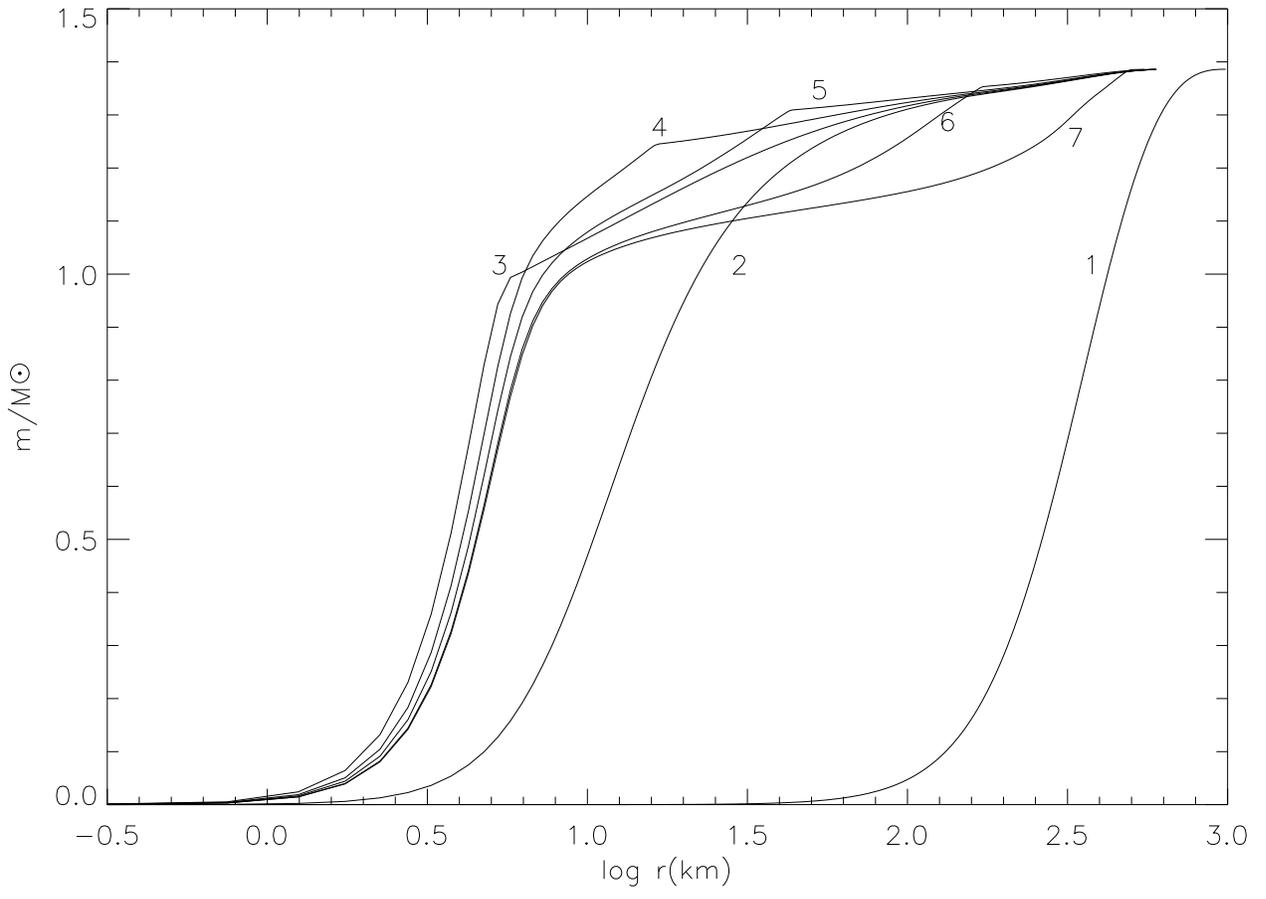
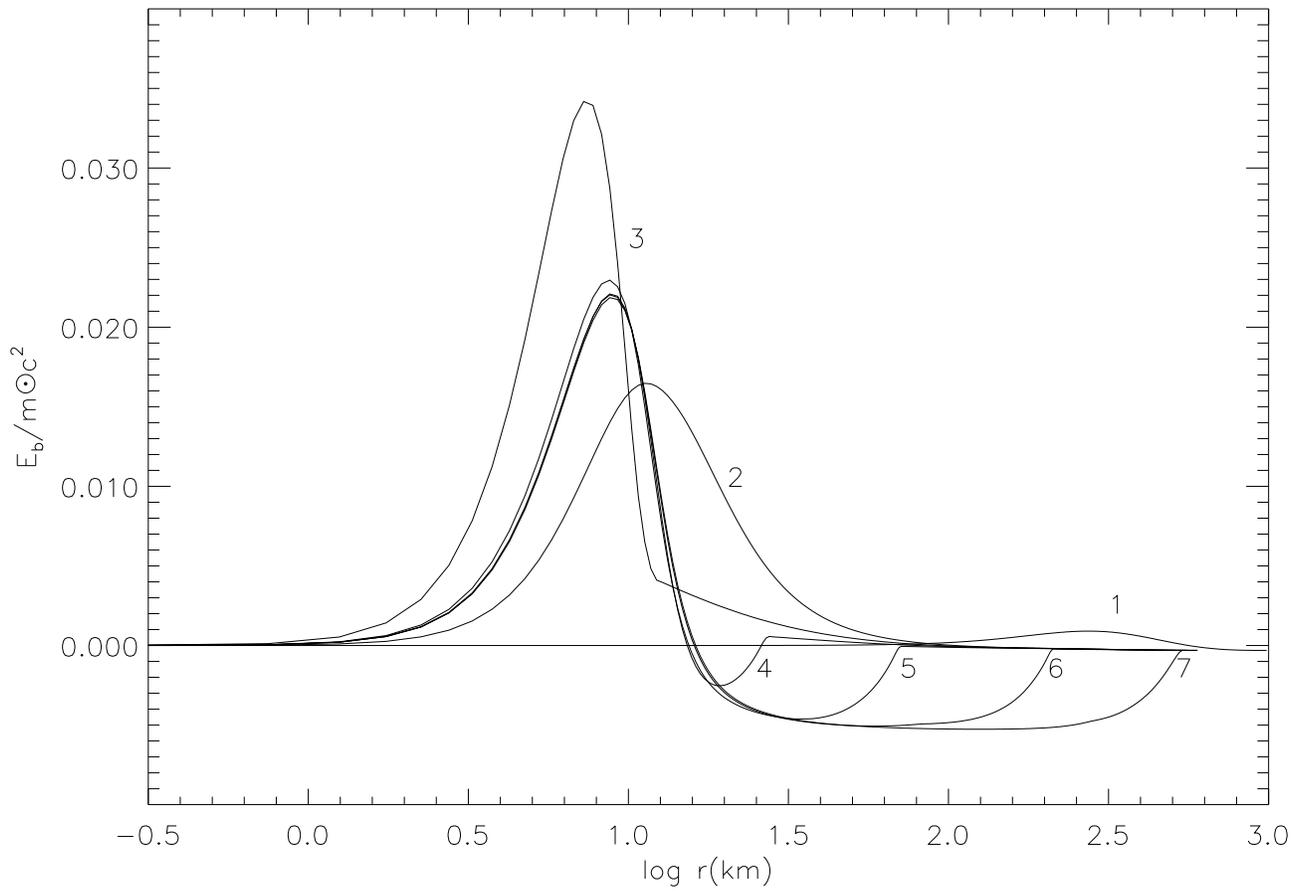
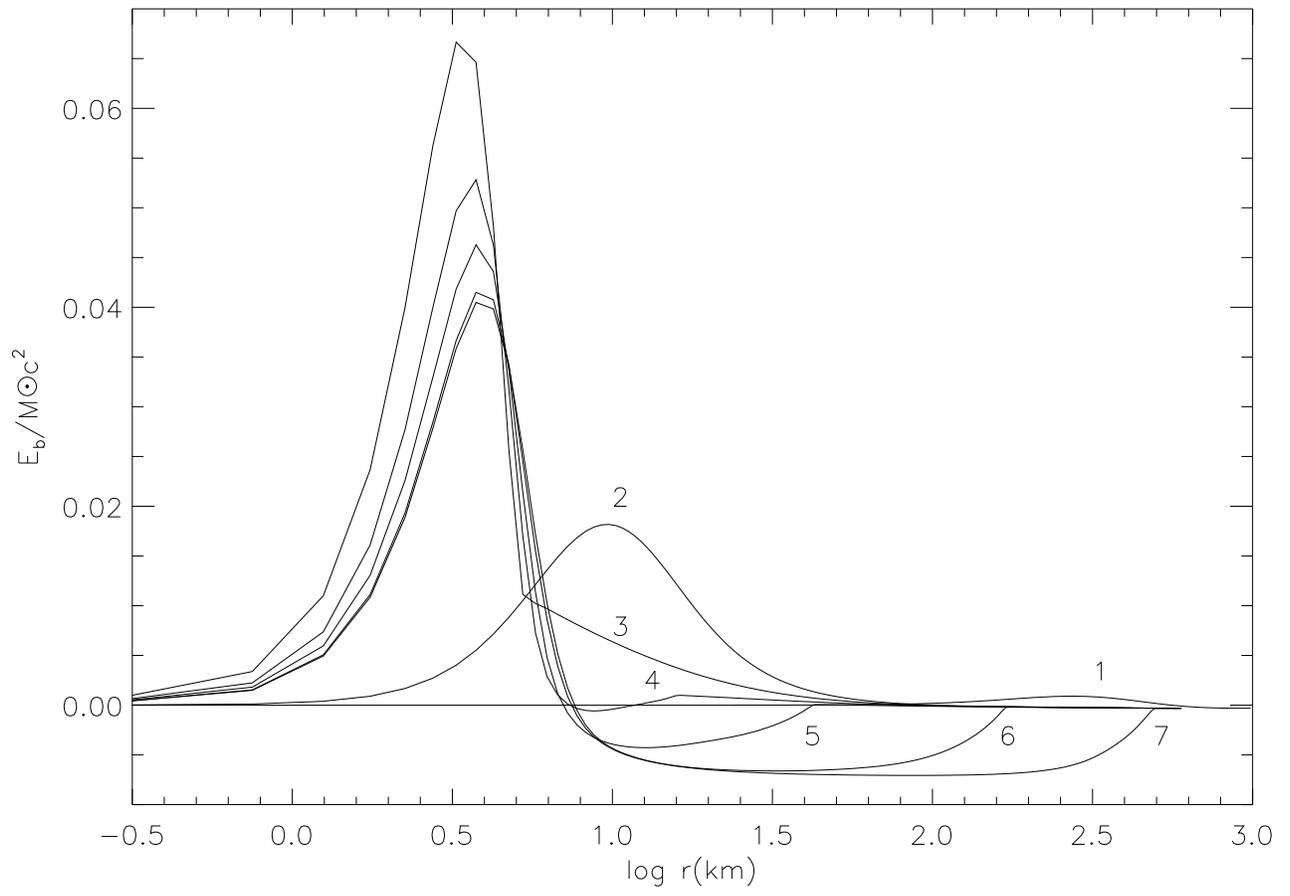
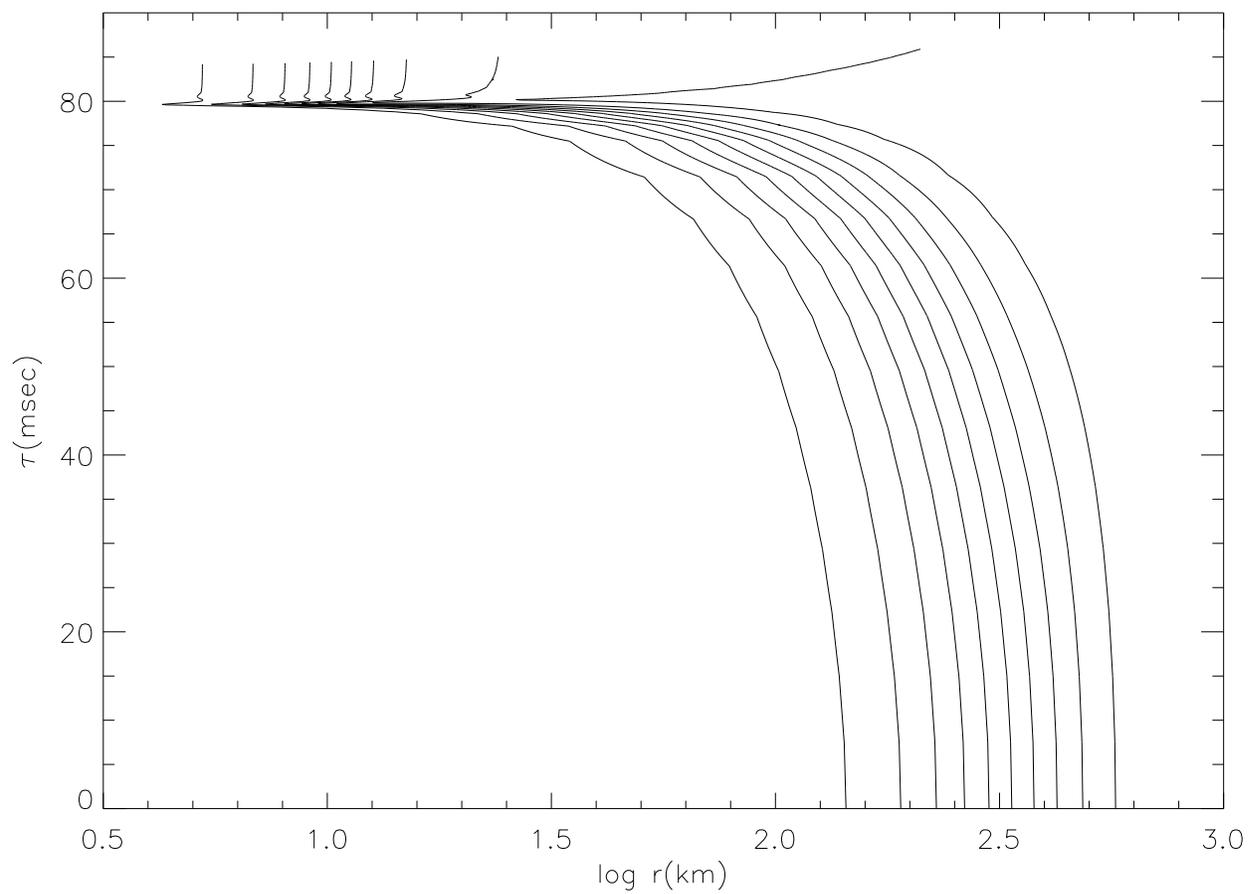
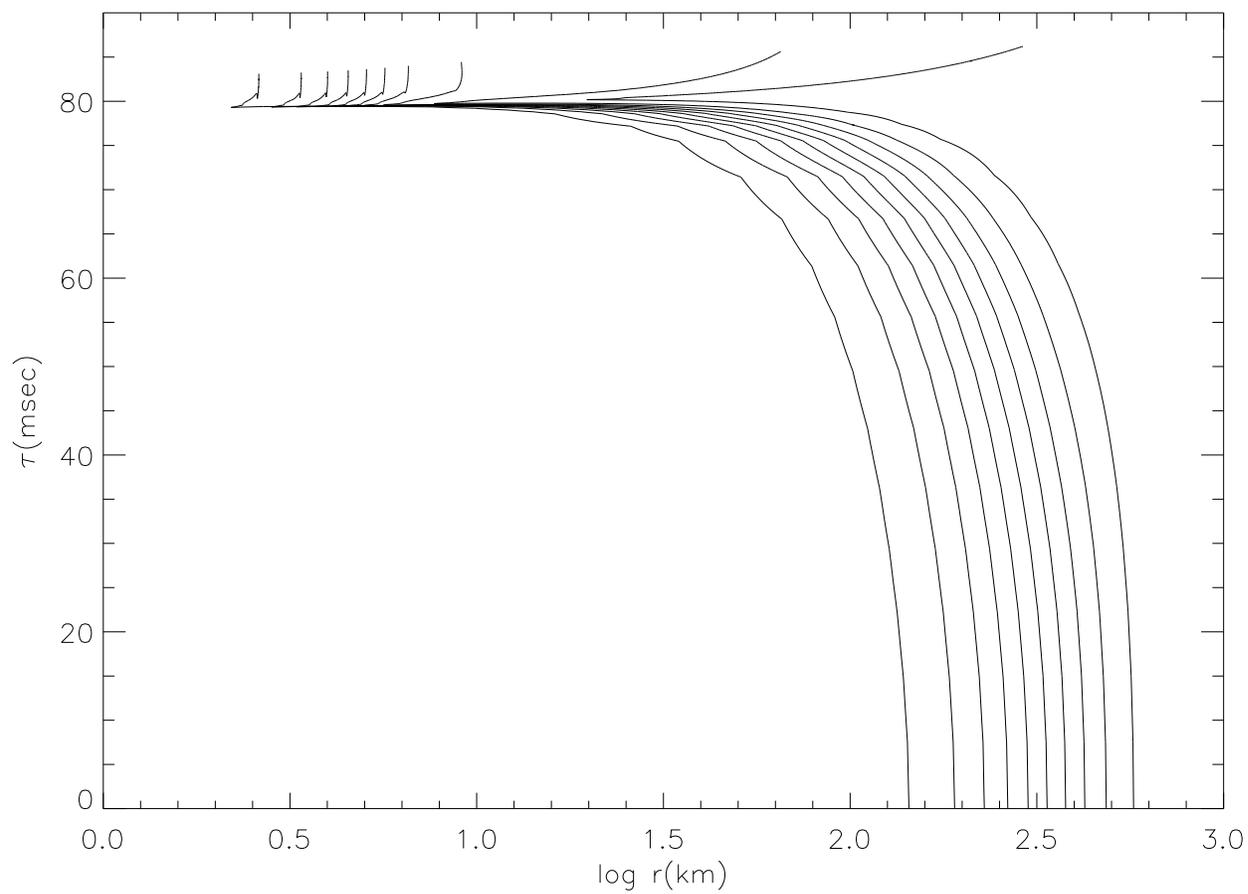
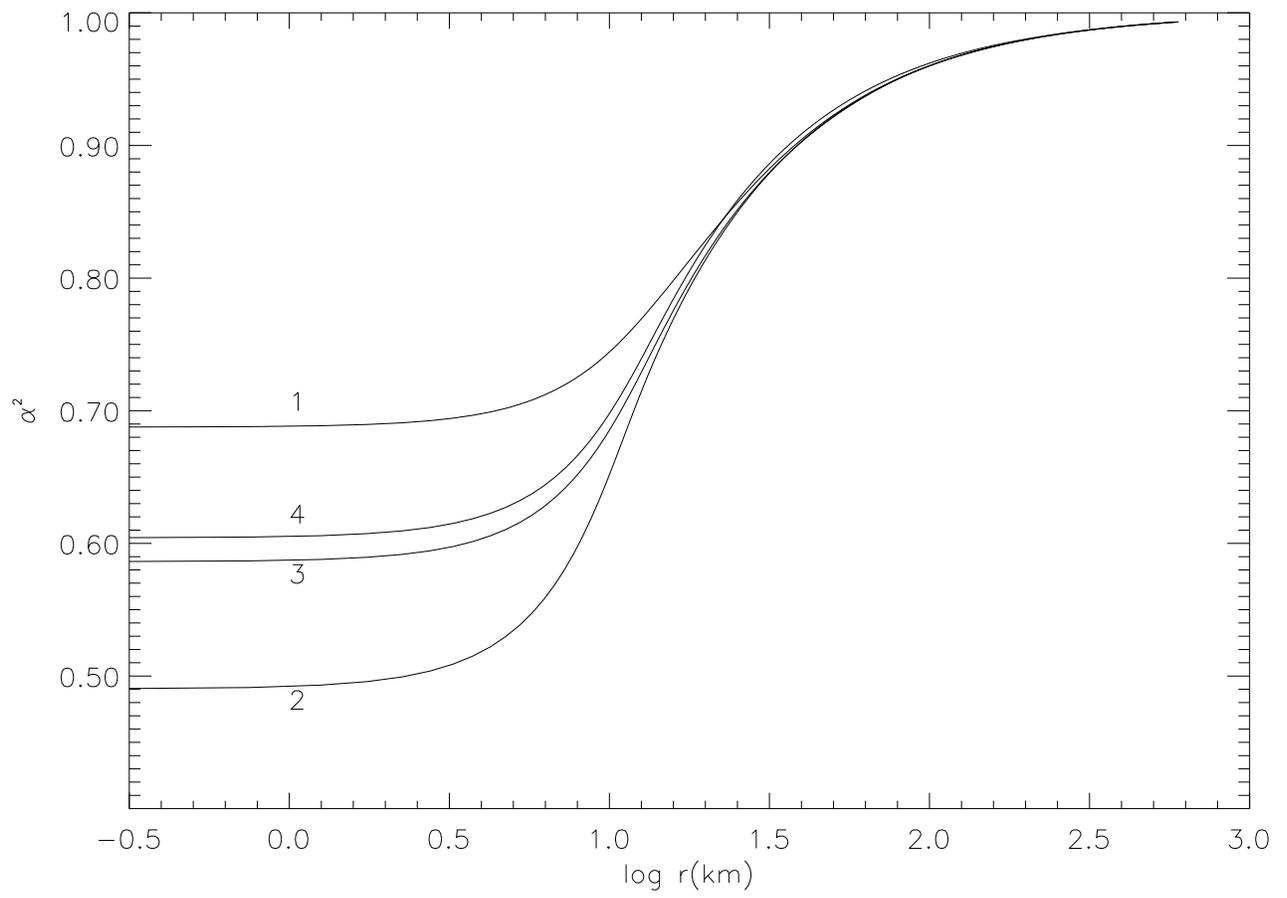
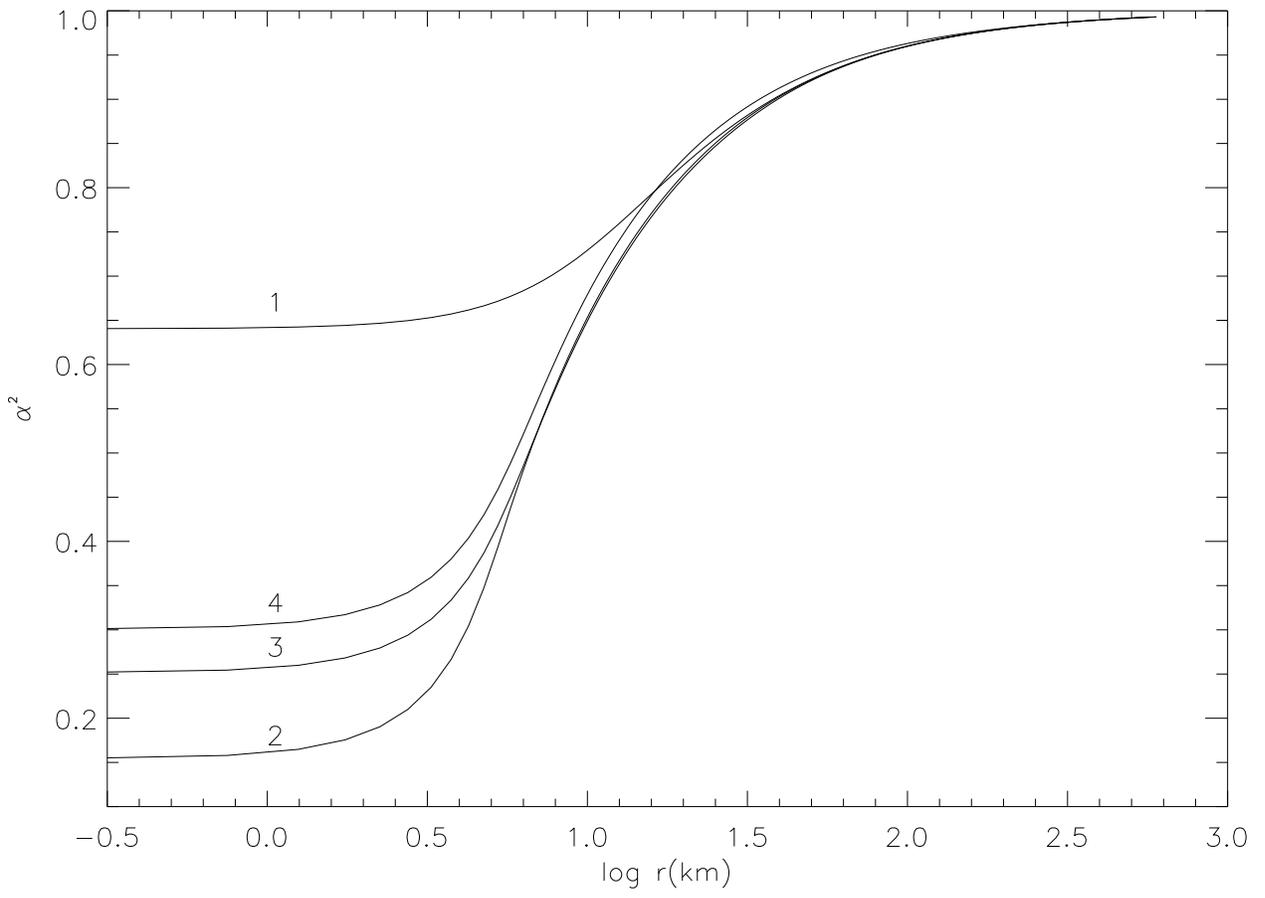
\end{document}